\documentclass[a4paper, 11pt, oneside]{article}
\pdfoutput=1
\usepackage{jheppub}

\usepackage{amssymb}
\usepackage{epsfig}
\usepackage{graphicx}% Include figure files
\usepackage{subfigure}
\usepackage{dcolumn}% Align table columns on decimal point
\usepackage{bm}% bold math
\usepackage[usenames ,dvipsnames]{xcolor}
\usepackage{slashed}
\usepackage{tikz}
\usetikzlibrary{positioning}
\tikzset{>=stealth}
\usepackage{epstopdf}
\usepackage{tikz-feynman}
\tikzfeynmanset{compat=1.0.0}
\usepackage[latin1]{inputenc}
\usetikzlibrary{trees}
\usetikzlibrary{decorations.pathmorphing}
\usetikzlibrary{decorations.markings}
\usepackage{verbatim}
\usepackage{subfigure}
\usepackage{xcolor}
\newcommand{\nc}{\newcommand}
\nc{\beq}{\begin{equation}}  \nc{\eeq}{\end{equation}}
\nc{\bea}{\begin{eqnarray}}  \nc{\eea}{\end{eqnarray}}
\nc{\baa}{\begin{array}}     \nc{\eaa}{\end{array}}
\nc{\bit}{\begin{itemize}}   \nc{\eit}{\end{itemize}}
\nc{\ben}{\begin{enumerate}} \nc{\een}{\end{enumerate}}
\nc{\bce}{\begin{center}}    \nc{\ece}{\end{center}}
\nc{\bpm}{\begin{pmatrix}}   \nc{\epm}{\end{pmatrix}}
\nc{\bvt}{\begin{verbatim}}  \nc{\evt}{\end{verbatim}}

\begin{document}

%\title{Leading-Order One-Loop Contribution to Dark-Matter-Nucleon Interaction in a Scalar Dark Matter Model with Tree-Level Cancellation Mechanism}
\title{Filtered pseudo-scalar dark matter and gravitational waves from first order  phase transition}

\author[1]{Wei Chao,}
\author[1]{Xiu-Fei Li,}
\author[2]{Lei Wang}

\affiliation[1]{Center for Advanced Quantum Studies, Department of Physics, Beijing Normal University, Beijing, 100875, China}
\affiliation[2]{Department of Physics, Yantai University, Yantai 264005,  China}
\emailAdd{chaowei@bnu.edu.cn}
\emailAdd{xiufeili@mail.bnu.edu.cn}
\emailAdd{leiwang@ytu.edu.cn}

\date{\today}
\abstract{
If dark matter (DM) acquires mass during a first order  phase transition,  there will be a filtering-out effect when DM enters the expanding bubble. In this paper we study the filtering-out effect for a pseudo-scalar DM, whose mass may partially come from a first order phase transition in the hidden sector.  We calculate the ratio of DM that may enter the bubble for various bubble wall velocities as well as various status of DM (in the thermal equilibrium, or out of the thermal equilibrium)  at the time of phase transition, which results in small penetration rate that may affect the final relic abundance of the DM. We further study the stochastic gravitational wave signals emitted by the hidden sector phase transition at the space-based interferometer experiments as the smoking-gun of this model. 
}

%\pacs{95.35.+d, 13.85.Tp, 14.80.-j, 98.70.Sa,}
%\keywords{Dark Matter, Cosmic Rays, AMS-02 Experiment}
\maketitle

%%%%%%%%%%%%%%%%%%%%%%%%%%%%%%%%%%%%%%%%%%%%%%%%%%%%%%%%%%%%%%%%%%%%%%%%%%%%%%%%%%%%%%%%%
\section{Introduction}
\label{s1}

Astrophysical observations show that about the 26.8\% of our Universe is made by  dark matter (DM)~\cite{Aghanim:2018eyx}, which cannot be addressed by the minimal Standard Model (SM) of particle physics.  What is the mass, spin, interactions as well as the thermal history of DM  has long been a puzzle for physicists,  which elicits competition for model buildings of DM  with mass ranging from $10^{-22}~{\rm eV}$ to $10^{55}~{\rm GeV}$. Many experiments are designed to directly probe DM by measuring the  recoil energy induced by the elastic scattering of DM with the target in underground laboratories, or indirectly probe DM by detecting the flux of a secondary cosmic rays injected by the decay or annihilation of DM. 

Among various DM models, the weakly interacting massive particles (WIMPs)~\cite{Goldberg:1983nd,Ellis:1983ew,Jungman:1995df,Servant:2002aq,Cheng:2002ej,Bertone:2004pz}, which explain the relic abundance of DM via the thermal freeze-out mechanism,  are well-motivated since they can address the observed relic density with DM mass at the electroweak scale and coupling as weak as the weak nuclear force.   WIMPs have been the main target of many DM direct detection experiments. However, no convincing  signal has been observed in any experiment,  which provides strong constraint on the WIMP-nucleon cross section with mass ranging from $\rm GeV$ to $\rm TeV$. For WIMP mass above  the $\rm TeV$ scale, the constraint of direct detection turns to be weak due to a suppressed DM number density.  However, there is an upper bound on the WIMP mass at about 100~$\rm TeV$~\cite{Griest:1989wd,Baldes:2017gzw,Smirnov:2019ngs}, above which  the WIMP interaction required by the observed relic density violates unitarity.  Several attempts beyond the traditional  freeze-out mechanism have been made to evade the constraint of unitarity, including the thermal freeze-in mechanism~\cite{Kolb:2017jvz},  the inelastic scattering against the particle in the thermal bath~\cite{Kim:2019udq}, the filtering-out effect during the first order phase transition~\cite{Baker:2019ndr,Chway:2019kft}, etc. 

The interplay of DM with cosmological phase transitions has been vastly investigated~\cite{Baker:2019ndr,Chway:2019kft,Chung:2011hv,Heurtier:2019beu,Hambye:2018qjv,Baker:2016xzo,Baker:2017zwx,Bian:2018mkl,Cohen:2008nb,Petraki:2011mv,Hall:2019rld,Chowdhury:2011ga,Fairbairn:2013uta,Chao:2017vrq,Chao:2015uoa,Ghorbani:2017jls,Alanne:2014bra,Hong:2020est,Marfatia:2020bcs,Schramm:1984bt,Chung:2011it,Baker:2018vos,Falkowski:2012fb,Huang:2017kzu,Bai:2018dxf,Shu:2006mm,Baldes:2017rcu,Gu:2017rzz,Gonderinger:2009jp,Carena:2011jy,Borah:2012pu,Gil:2012ya,Ahriche:2013zwa,Ghorbani:2019itr},  which shows that both the freeze-out temperature and the number density of DM can be tightly correlated with the  phase transition if a DM mass originates from the spontaneous symmetry breaking.   In this paper, we follow the idea of Refs.~\cite{Baker:2019ndr,Chway:2019kft} and investigate the filtering-out  effect of a pseudo-scalar DM passing through an expanding bubble by assuming that the mass of a pseudo-scalar DM only partially originates from the phase transition. Compared with previous studies, this work clarifies the following respects: 
\begin{itemize}
\item We derive the filtering-out condition for a DM, whose mass only partially originates from the phase transition, in Eq.~(\ref{condition}), which is slightly different from that of the fermionic DM.  
\item 
We calculate the penetration rate and the DM relic abundance for various bubble wall velocities and various DM status at the time of phase transition. More especially, we study the filtering-out effect for a case where the pseudo-scalar DM freezes out before the phase transition, which has not been considered before.  
\end{itemize}
We further study the stochastic gravitational wave (GW) signal induced by the first order phase transition at the space-based interferometer as a smoking gun for this scenario.  Our result shows that the GW energy spectrum is reachable by the LISA,  Taiji, TianQin, BBO, DECIGO and Ultimate-DECIGO experiments.

The remaining of the paper is organized as follows: In section II we present the model in detail. Section III is devoted to the calculation of filtering-out effect of the pseudo-scalar DM. Section IV is focused on the investigation of GW signals at the space-based interferometer. The last part is concluding remarks.

%%%%%%%%%%%%%%%%%%%%%%%%%%%%%%%%%%%%%%%%%%%%%%%%%%%%%%%%%%%%%%%%%%%%%%%%%%%%%%%%%%
\section{The model}\label{Sec_model}
We begin our discussion by specifying a pseudo-scalar DM model, which extends the SM with an extra complex scalar $S$ and a softly broken global $U(1)$ symmetry.  The CP-odd component of $S$ is taken as the DM candidate stabilized by the CP symmetry, $S\leftrightarrow S^*$.  The tree-level potential can be written as
\begin{eqnarray}
\label{potential1}
V(H,S) &=& -\mu_h^2 |H|^2 -\mu_s^2 |S|^2 + \lambda_h |H|^4 + \lambda_s |S|^4 + \lambda_{sh} |H|^2 |S|^2 \nonumber \\
&& - \frac{\mu_2^2}{2}\left(S^2 + {S^\star}^2 \right)\ + \frac{\mu_3}{2}\left(S^3 + {S^\star}^3 \right)+ \frac{\mu_4}{2}\left(S^4 + {S^\star}^4 \right),
\end{eqnarray}
where  $\mu_2$ and $\mu_3$ are real parameters with mass dimension, $\mu_4$ is a real dimensionless coupling, $H$ is the SM Higgs. $S$  takes the following form 
\begin{eqnarray}
S=\frac{v_s+s+i\chi}{\sqrt{2}},
\end{eqnarray}
where $v_s$ is the vacuum expectation value of $S$. Eq.~(\ref{potential1}) is invariant under the CP transformation $S \to S^*$, therefore $\chi$ is a DM candidate.  Applying the minimization condition, $v_s$ can be written as
\begin{eqnarray}
v_s=\frac{-3\mu_3+\sqrt{32\lambda_s\mu_2^2+9\mu_3^2+32\lambda_s\mu_s^2}}{4\sqrt{2}\lambda_s} \; ,
\end{eqnarray}
and the mass eigenvalues of $s$ and $\chi$ are
\begin{align}
%m_s^{}&=&\sqrt{ 2\lambda_s v_s^2 +\frac{3}{2\sqrt{2}}\mu_3 v_s + 2\mu_4 v_s^2 } \\ 
m_s^{}=\sqrt{ 2\lambda_s v_s^2 +\frac{3}{2\sqrt{2}}\mu_3 v_s} \; , \hspace{1cm}
%m_\chi^{} &=& \sqrt{2\mu_2^2 -\frac{9}{2\sqrt{2}}\mu_3 v_s -4\mu_4 v_s^2} \label{mchi}
m_\chi^{} = \sqrt{2\mu_2^2 -\frac{9}{2\sqrt{2}}\mu_3 v_s}~~, \label{mchi}
\end{align}
where we have set $\mu_4=0$ for convenience. It should be noted that we have assumed that the coupling of the Higgs portal interaction, $\lambda_{sh}^{} (H^\dagger H) (S^\dagger S)$, where $H$ is the SM Higgs doublet,  is negligibly small for simplicity.  Since we are interested in the phase transition at a temperature above the electroweak scale, such a simplification is reasonable.  A  general Higgs potential including the Higgs portal interaction  as well as the minimization conditions, physical parameters and the effective potential at the finite temperature are given in the Appendix \ref{appA}.  For such a real case, the CP-even scalar $s$ will decay into SM particles due to the mixing with the SM Higgs and there are Higgs-portal interactions, $({1/4}) \lambda_{sh}^{} v_h^{}  \chi^2  h^2 $ and $(1/2) \lambda_{sh} \chi^2 h$,  with coupling strength proportional to $\lambda_{sh}$.

%%%%%%%%%%%%%%%%%%%%%%%%%%%%%%%%%%%%%%%%%%%%%%%%%%%%%%%%%%%%%%%%%%%%%%%%%%%%%%%%%%%%%%%%%%%%%%%%%%%%%%%%%%%%%%%%%
\section{Filtering-out effect} \label{Sec_Tree}
It has been shown~\cite{Baker:2019ndr,Chway:2019kft} that the filtering-out effect can be significant for a relativistic  fermionic DM entering an expanding bubble.  In our model, $\chi$ partially acquire mass from the  phase transition in the dark sector, which can be strongly first-order due to the existence of the cubic term at the tree-level in the potential.   We study the filtering-out effect  of $\chi$  for various bubble wall velocities and various status of DM (in thermal equilibrium or out of thermal equilibrium) at the time of phase transition in this section.

\subsection{Filtering-out condition for a massive particle}

A first-order phase transition is processed by the bubble nucleation of the true vacuum.  We assume the  bubble wall moves in the negative $z$ direction with velocity $v_w$, and let $F^{p}$ represents the plasma's rest frame, let $F^w$ represents the bubble wall's rest frame.  During the first order phase transition,  a massive particle $\chi$ that is hitting on the expanding bubble wall with momentum $p=(p_x,~p_y,~p_z)$ in the $F^p$ frame  is able to penetrate into the wall only when its  $z$-direction momentum in the $F^w$ frame is sufficiently large,
\begin{eqnarray}
\gamma_w (p_z + v_w E) > \sqrt{\Delta m^2} \; , \label{condition}
\end{eqnarray}
where $\gamma_w =1/\sqrt{1-v_w^2}$ is the Lorentz factor, $E=\sqrt{p^2+m_0^2}$ with $m_0$ the mass of $\chi$ outside the bubble, $\Delta m^2 = m_{\chi}^2 -m_0^2$ with $m_{\chi}^{}$ the mass of $\chi$ after phase transition.  Eq.~(\ref{condition}), which is derived from the energy and momentum conservation, serves as filtering-out condition for a massive particle. Compared with condition of massless case,  $\gamma_w(p_z+v_w |p|)>m_{\chi}$~~\cite{Baker:2019ndr} , where DM is always in thermal equilibrium outside the bubble,  the condition of  the massive case is more complicated, and a nonzero $m_0$ may lead to a different status of $\chi$ at the time of bubble nucleation, thus it is worth of study in detail.

\subsection{DM number density and penetration rate $R_\chi$}
In this subsection, we investigate the DM penetration rate $R_\chi$ (defined in Eq.~(\ref{ratio})) with different bubble velocity $v_w$. To do so, we first estimate the DM number density inside the bubble wall using the method given in Ref.\cite{Chway:2019kft,Marfatia:2020bcs,Hong:2020est}, then we use it to calculate the DM relic abundance. We fist assume DM is in the thermal equilibrium outside the bubble wall, such that its distribution function follows the Bose-Einstein distribution 
\begin{equation}
f_\chi=\frac{1}{e^{\gamma_w(\sqrt{|\tilde{\textbf{p}}|^2+m_0^{2}}-v_w\tilde{p}_z)/T_n}-1} \; ,
\end{equation}
%where $\mu_\chi$ is the chemical potential. 
%The number density of $s$ is given by
%\begin{equation}
%n_s^{\rm f.v.}=2\int \!\! \frac{d^3\textbf{p}}{(2\pi)^3}f_s^{\rm f.v.}(\textbf{p})\approx\frac{3\zeta(3)}{2\pi^2}T_n^3+\frac{\mu_\chi}{6}T_n^2.
%\end{equation}
where $T_n$ stands for the bubble nucleation temperature, $\tilde p_z$ is the momentum in the  wall's rest frame. Because of the energy conservation, only $\chi$ with $\tilde p_z^2 >\Delta m^2$ can enter the bubble.  The particle flux coming from the false vacuum per unit area and unit time can be written as\cite{Chway:2019kft,Hong:2020est}
\begin{align}
\tilde{J_\chi}&=g_\chi\int \!\! \frac{d^3\tilde{\textbf{p}}}{(2\pi)^3}\frac{\tilde p_z}{\tilde E} f_\chi\Theta \left(\tilde p_z-\sqrt{\Delta m_{}^2 }\right) \; ,\label{xxxx}
%%% \notag\\
%&=g_\chi\int \!\! \frac{d^3\textbf{p}}{(2\pi)^3}\frac{\gamma_w(p_z+v_w|\textbf{p}|)}{\gamma_{w}(|\textbf{p}|+v_wp_z)}\tilde f_\chi^{\rm f.v.}(\textbf{p})\Theta \left (\gamma_w(p_z+v_w|\textbf{p}|)-m_\chi^{\rm in}\right) 
%\notag\\
%&\approx g_\chi\int \!\! \frac{d^3\textbf{p}}{(2\pi)^3}\frac{p_z}{|\textbf{p}|}\tilde f_\chi^{\rm f.v.}(\textbf{p})\Theta(p_z-m_\chi^{\rm in}) 
%%%%%%%%%%%%%%%%%%%%%%%%%%%%%%%%%%%%%%%%%%%%%%%%%%%%%%%%%%%%%%%%%%%%%%%%%%%%%%%%%%%%%%%%%%%%%%%%%%%%%%
%%% &= g_\chi\int \!\! \frac{d^3\tilde{\textbf{p}}}{(2\pi)^3}\frac{\tilde p_z}{|\tilde{\textbf{p}}|}\tilde f_\chi(\tilde{\textbf{p}})\Theta \left(\tilde p_z-\delta m_\chi \right)
%%% \notag\\
%%% &= g_\chi\int \!\! \frac{d\tilde p_\perp d\tilde p_z}{(2\pi)^2}\frac{\tilde p_\perp \tilde p_z}{\sqrt{\tilde p_\perp^2+\tilde p_z^2+m_0^2}}\tilde f_\chi(\tilde{\textbf{p}})\Theta \left(\tilde p_z-\delta m_\chi \right)
%%% \notag\\
%%% &= g_\chi\int \!\! \frac{d\tilde p_\perp d\tilde p_z}{(2\pi)^2}\frac{\tilde p_\perp \tilde p_z}{\sqrt{\tilde p_\perp^2+\tilde p_z^2+m_0^2}}\frac{1}{e^{\gamma_w\sqrt{\tilde p_\perp^2+\tilde p_z^2+m_0^2}/T_n}}e^{\gamma_wv_w\tilde p_z/T_n}\Theta \left(\tilde p_z-\delta m_\chi \right)
%%% \notag\\
%%% &= \frac{g_\chi T_n}{\gamma_w}\int \!\! \frac{d\tilde p_z}{(2\pi)^2}\tilde p_ze^{-\left(\sqrt {\tilde p_z^2+m_0^2}-v_w\tilde p_z \right)\gamma_w/T_n}\Theta \left(\tilde p_z-\delta m_\chi \right)
\end{align}
where $g_\chi=1$  being the DM degrees of freedom. Then the DM number density inside the bubble $n_\chi^{\rm in}$ in plasma rest frame can be written as~\cite{Chway:2019kft,Hong:2020est}
\begin{align}
n_\chi^{\rm in}&=\frac{\tilde{J_\chi}}{\gamma_{w}v_w} \; .
\label{number density}
\end{align}
The DM penetration rate $R_\chi$ can be defined as 
\begin{equation}
\label{ratio}
R_\chi=\frac{n_\chi^{\rm in}}{n_\chi^{\rm out}} \; ,
\end{equation}
where $n_\chi^{\rm out}$ is the DM number density outside the bubble wall at the time of the phase transition.  In the following, we will calculate this ratio for cases where DM is either in the thermal equilibrium or out of the thermal equilibrium at the time of phase transition, and estimate the impact of various bubble wall velocities to the penetrate rate.

\begin{figure}[t]
	%\label{annihilation}[htbp]
	%	\subfigure [dffdgxcd]
	\centering
	\begin{tikzpicture}
	\begin{feynman}
	\vertex (i1);
	\vertex [below right=of i1] (i2);
	\vertex [below left=of i2] (i3);
	\vertex [above right=of i2] (i4);
	\vertex [below right=of i2] (i5);
	
	\diagram* {
		(i1) -- [scalar, ultra thick] 
		(i2) -- [scalar, ultra thick] (i3),
		(i4) -- [scalar, ultra thick] (i2) -- [scalar, ultra thick] (i5),
	};
	\end{feynman}
	\node[black, thick] at (-0.2,0.5) {$\chi$};
	\node[black, thick] at (-0.2,-2.5) {$\chi$};
	\node[black, thick] at (2.5,0.5) {$S$};
	\node[black, thick] at (2.5,-2.5) {$S$};
	\end{tikzpicture}
	\hspace{0.5cm}
	\begin{tikzpicture}
	\begin{feynman}
	\vertex (i1);
	\vertex [below=2cm of i1] (i2);
	\vertex [right=2cm of i1] (i3);
	\vertex [right=2cm of i2] (i4);
	\vertex [right=2cm of i3] (i5);
	\vertex [right=2cm of i4] (i6);
	
	\diagram* {
		(i1) -- [scalar, ultra thick] (i3)-- [scalar, ultra thick] (i4),
		(i2) -- [scalar, ultra thick] (i4),
		(i3) -- [scalar, ultra thick] (i5),
		(i4) -- [scalar, ultra thick] (i6),
		
	};
	\end{feynman}
	\node[black, thick] at (1,0.5) {$\chi$};
	\node[black, thick] at (1,-2.5) {$\chi$};
	\node[black, thick] at (2.3,-1) {$\chi$};
	\node[black, thick] at (3,0.5) {$S$};
	\node[black, thick] at (3,-2.5) {$S$};
	\end{tikzpicture}
	\hspace{0.5cm}
	\begin{tikzpicture}
	\begin{feynman}
	\vertex (i1);
	\vertex [below=2cm of i1] (i2);
	\vertex [right=2cm of i1] (i3);
	\vertex [right=2cm of i2] (i4);
	\vertex [right=2cm of i3] (i5);
	\vertex [right=2cm of i4] (i6);
	
	\diagram* {
		(i1) -- [scalar, ultra thick] (i3) -- [scalar, ultra thick] (i4),
		(i2) -- [scalar, ultra thick] (i4),
		(i3) -- [scalar, ultra thick] (i6),
		(i4) -- [scalar, ultra thick] (i5),	
	};
	\end{feynman}
	\node[black, thick] at (1,0.5) {$\chi$};
	\node[black, thick] at (1,-2.5) {$\chi$};
	\node[black, thick] at (2.3,-1) {$\chi$};
	\node[black, thick] at (4,0.5) {$S$};
	\node[black, thick] at (4,-2.5) {$S$};
	\end{tikzpicture}
	\caption{\small  Annihilation channels for $\chi\chi\leftrightarrow SS$.}
	\label{annihilation}
\end{figure}
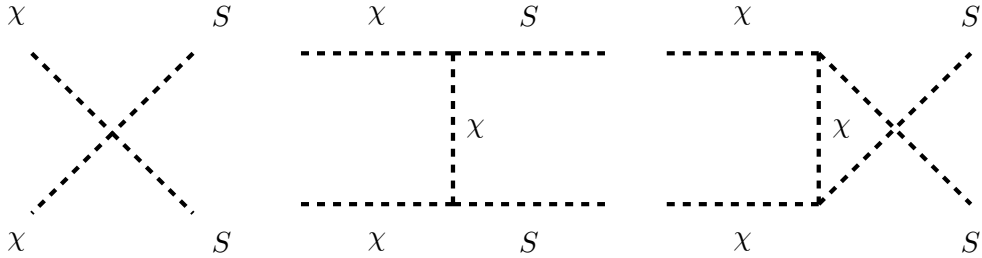

%\subsection{Freeze out styles and relic density}
\subsection{Scenario A: DM freeze-out after/during the phase transition}

In this subsection we calculate the penetration rate for cases where DM freezes out during the phase transition or after the phase transition.  The first case is the so-called  spontaneous freeze-out mechanism~\cite{Heurtier:2019beu}.   For convenience, it is necessary to define two variables
\begin{eqnarray}
Y_\chi=\frac{n_\chi}{s},~~x=\frac{m_\chi}{T},
\end{eqnarray}
where $s$ is the entropy density $s=2\pi^{2}g_{* S} T^3/45$ and $g_{* S} \sim 106.75$ is the number of relativistic degrees of freedom.  Without taking into account the filtering-out effect, the evolution of the number density of DM is governed by the Boltzmann equation, 
\begin{eqnarray}
{dY_\chi\over dx} = - {s\langle\sigma v\rangle\over H x}
\left(Y_{\chi}^2 - Y_{\rm eq}^2\right),
\label{beq}
\end{eqnarray}
where
\begin{eqnarray}
	Y_{\rm eq} = {45\over 4\pi^4}{g_{\chi}\over
	g_{*S}} x^2 K_2(x) 
	\end{eqnarray}
and $K_2(x)$ is the modified Bessel function of the order $2$,  $H= 1.66g_{* S}^{1/2}T^2 /M_{\rm pl}$  is the Hubble constant with $M_{\rm pl}=1.22\times 10^{19}~{\rm GeV}$ being the Planck mass, $\langle \sigma v\rangle$ is the thermal average of the reduced annihilation cross section with the annihilation processes given in the Fig.~\ref{annihilation}~\footnote{We have neglected the annihilation into SM Higgs for simplification. }.  For a real estimation, we need to calculate the relic abundance of DM in three steps: 
\begin{itemize}
\item Step-I: Solving the Eq.~(\ref{beq}) for $T>T_n$, where $T_n$ is the bubble nucleation temperature;
\item Step-II: Evaluating the filtering-out effect induced by the first order phase transition and match $Y_\chi$ outside the bubble to the value inside the bubble;
\item Step-III: solving the  Eq.~(\ref{beq}) for $T<T_n$ to get the final relic abundance.
\end{itemize}
Without loss of generality, we choose the bubble wall velocity as $v_w=0.4$\footnote{Even if $v_w=0.4$, there is still a very strong  filtering-out effect. We will discuss the effect of different $v_w$ on DM filtering-out effect in the section~\ref{3.4}.} and the bubble nucleation temperature $ T_{n}= 150$ GeV which is  close to the electroweak phase transition temperature. Here we fix $-\mu_3= v_s= 2m_s$  and  take $-\mu_3$ as  a free parameter. We consider two scenarios: $m_0=0$ and $m_0=2~{\rm TeV}$, in both of which DM is in thermal equilibrium with the thermal bath at the time phase transition.

For $m_0=0$ TeV, DM is massless outside the bubble and is kept in thermal equilibrium through the annihilation process $\chi\chi\leftrightarrow SS$ (See Fig.~\ref{annihilation}). When the Universe cools  down to phase transition temperature $T_n$, DM with sufficient energy can penetrate the bubble wall and acquires mass through the Higgs mechanism. 
As a result, the reaction $\chi\chi\leftrightarrow SS$ can be out of equilibrium abruptly after the phase transition due to the reduction of number density, which is quite similar to the spontaneous freeze-out mechanism~\cite{Heurtier:2019beu}. 
 % where DM is relativistic and $Y$ is a constant outside the bubble. 
 \begin{figure}[htbp]
 	\centering
 	\includegraphics[height=2.7in,angle=0]{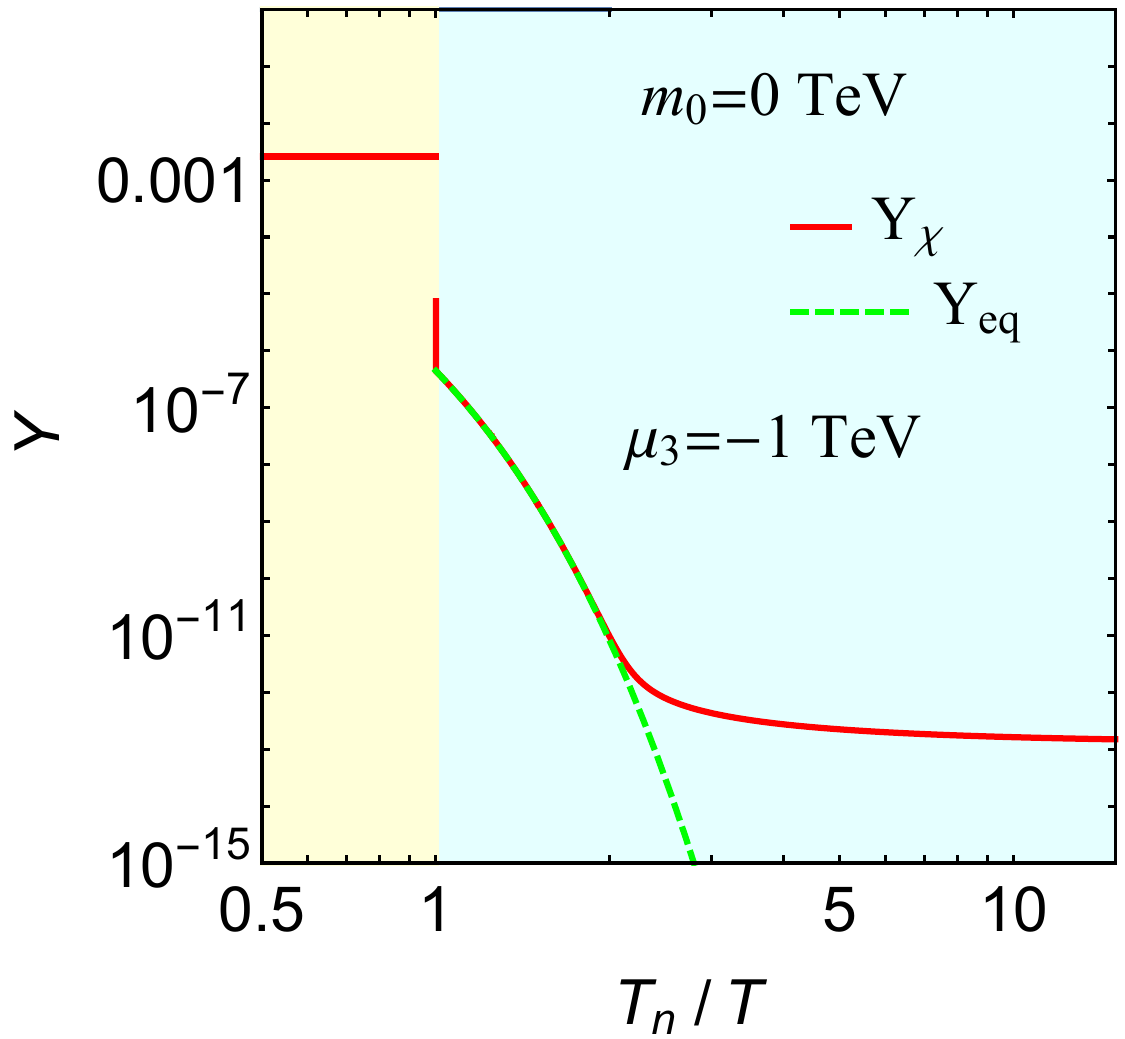}
 	\includegraphics[height=2.7in,angle=0]{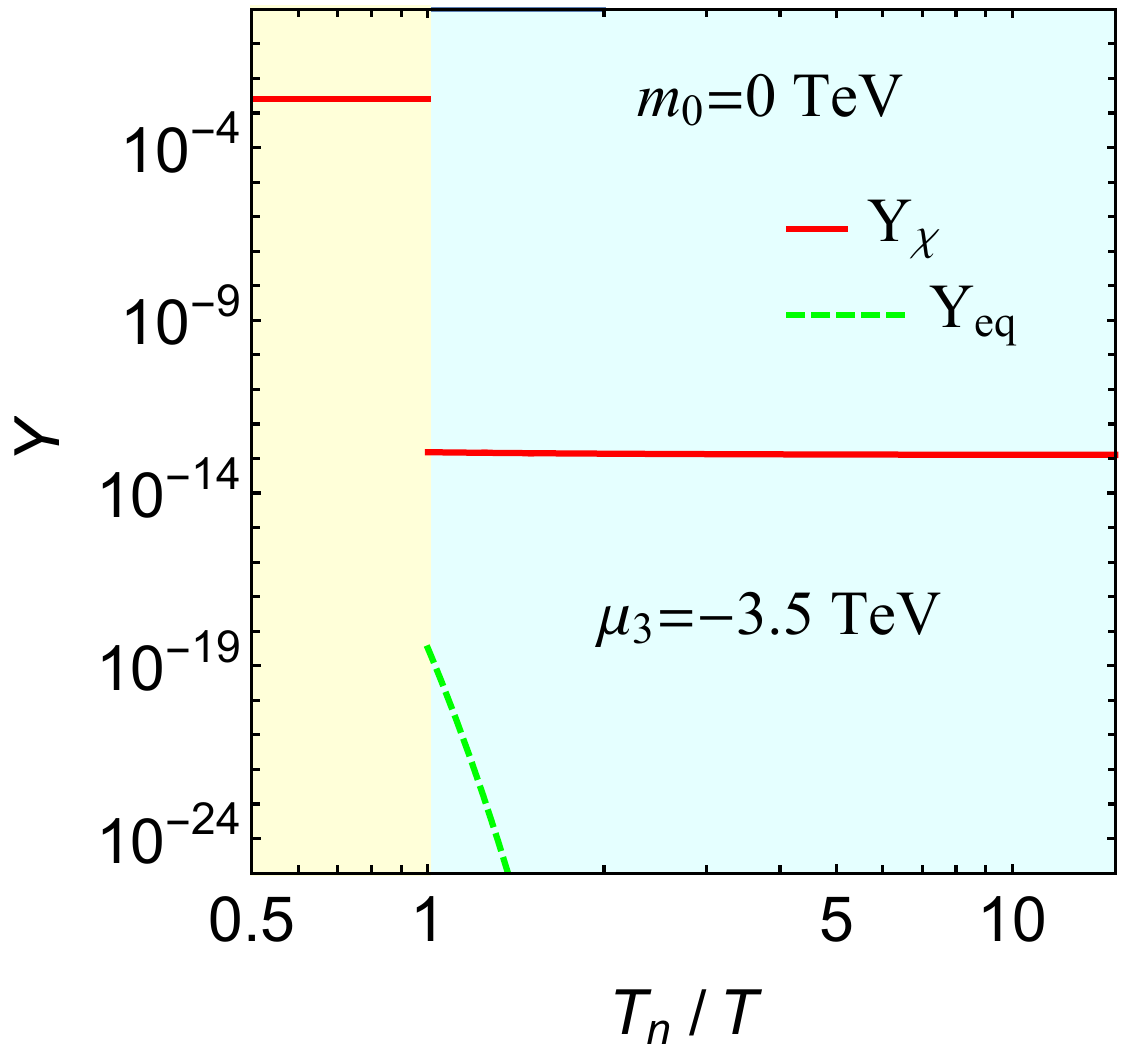} \\
	\includegraphics[height=2.7in,angle=0]{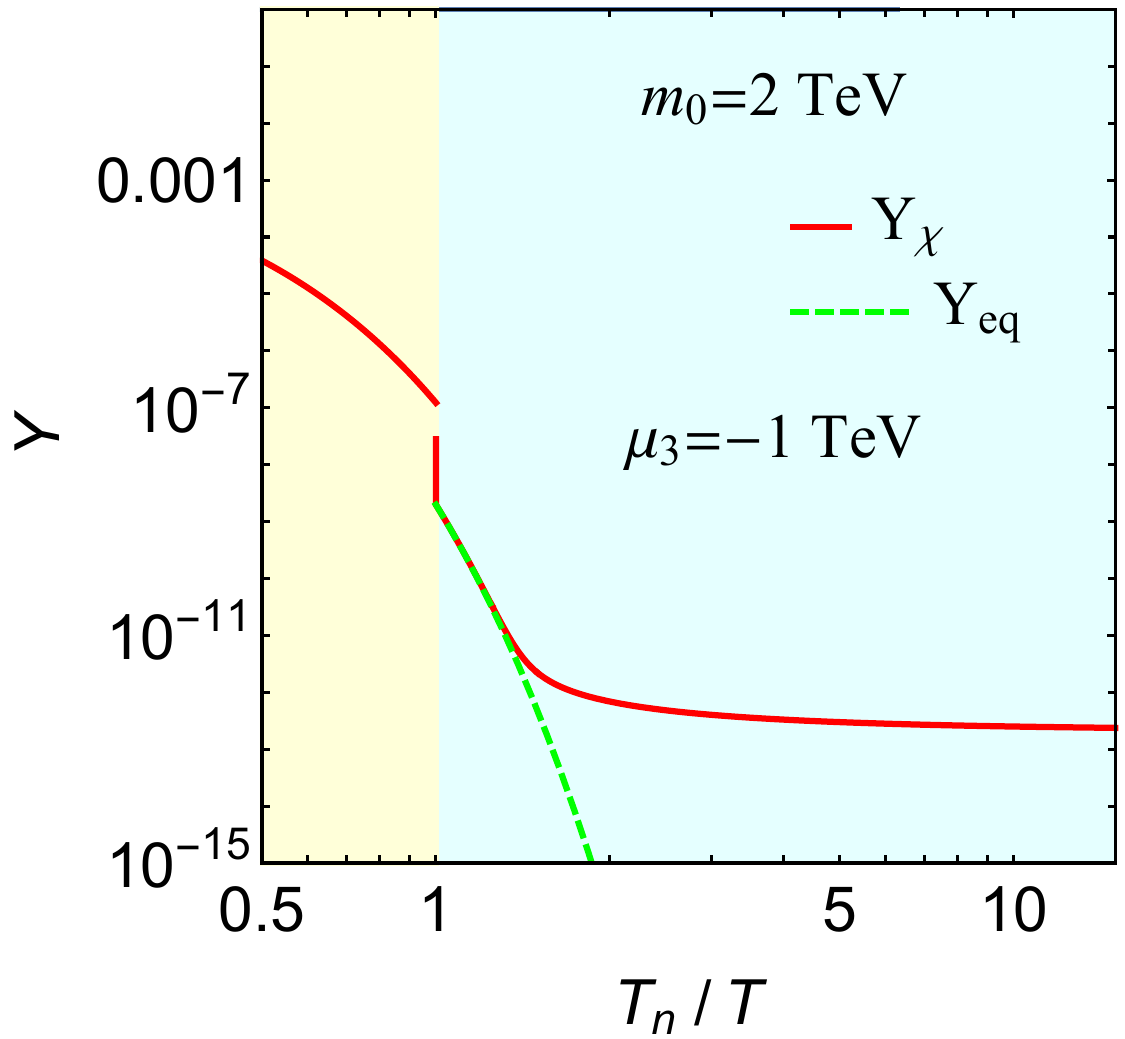}
	\includegraphics[height=2.7in,angle=0]{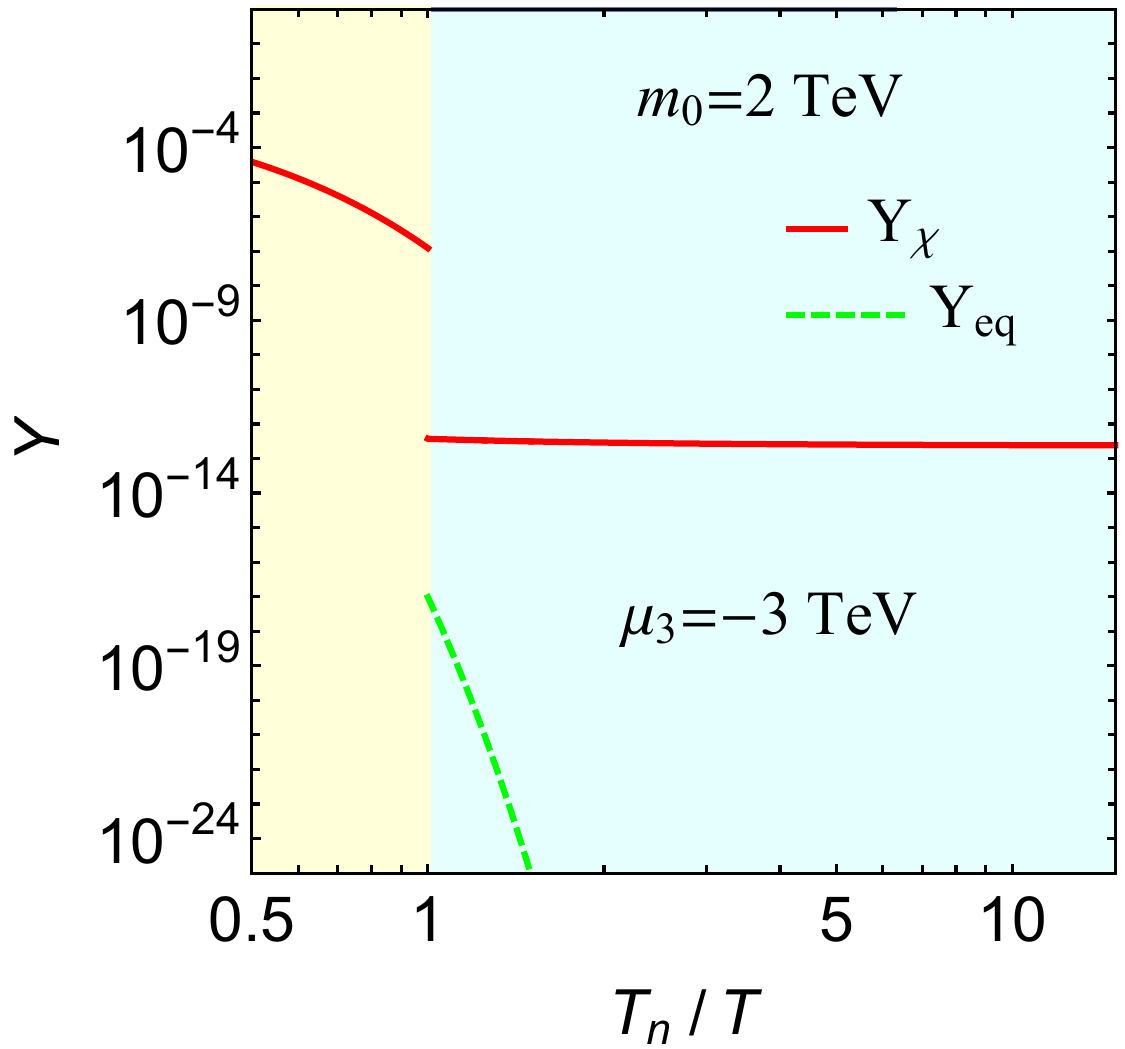}
 	\caption{\small \label{1types}
 	DM freeze-out after/during the phase transition. The red line stands for $Y_{\chi}$, green line is the  $Y_{\rm eq}$. The yellow(cyan) regime represents the case of outside(inside) the bubble, where the bubble nucleation temperature is located at $T_n =150~ {\rm GeV}.$
 	}
 \label{M0}
 \end{figure}
Fig.~\ref{1types} shows $Y_\chi$ as the function of $T_n/T$ for $m_0=0$ and $2$ TeV, where $T_n$ is the bubble nucleation temperature.
Plots in the top-left and top-right ($m_0=0~{\rm TeV}$) correspond to  $\mu_3=-1~{\rm TeV}$ and $-3.5~{\rm TeV}$, respectively. For the case of $\mu_3=-1~{\rm TeV}$ (left-panel), the DM penetrates the bubble wall at $x\sim 12$ and the penetration rate is $R_\chi\sim5.04\times10^{-3}$. DM number density $n_{\chi}^{\rm in}$ at $T=150$ GeV is much smaller than that of outside the bubble $n_{\rm \chi}^{\rm out}\sim T^3$, due to the very strong bubble filtering-out effect. Nevertheless, $n_{\chi}^{\rm in}$ is still higher than the  number density  required by thermal equilibrium $n_{\chi}^{\rm eq}\sim(m_\chi T)^{2/3} e^{-m_\chi/T}$.  The DM number density will experience a sharp decline to the value of thermal equilibrium via the reaction $\chi \chi \to S S$.  For the case of  $-\mu_3= 3.5 $ TeV (plot in the top-right),  DM penetrates the bubble wall at $x\sim42$ and  the penetration rate  is $R_\chi\sim8.14\times10^{-11}$. The DM turns to be very heavy and its number density is very small, so it freezes out immediately after entering the bubble, where $\frac{\Gamma(x\sim 42)}{H(x\sim 42)}\sim 0.20$.  Plots in the bottom-left and bottom-right of the Fig.~\ref{1types} show the filtering-out effect for  $m_{0}= 2$ TeV where DM is non-relativistic, but is still in thermal equilibrium at the time of phase transition.  In the bottom-left plot, $-\mu_3= 1$ TeV, the DM penetrates the bubble wall at $x=18$ and the penetration rate is $R_\chi\sim2.29\times10^{-1}$. %This value is smilar with the case of $m_0=0$. 
After a rapid  decline in $n_{\chi}^{\rm in}$, DM can achieve thermal equilibrium, which is the same as the case of $m_0=0$ TeV. Plot in the bottom-right  has  $-\mu_3= 3$ TeV,  DM can freeze out directly at $T=150$ GeV where $\frac{\Gamma(x\sim 38)}{H(x\sim 38)}\sim 0.56$ and the penetration rate is $R_\chi\sim3.04\times10^{-6}$. 

\begin{figure}[htbp]
	\centering
	\includegraphics[height=2.7in,angle=0]{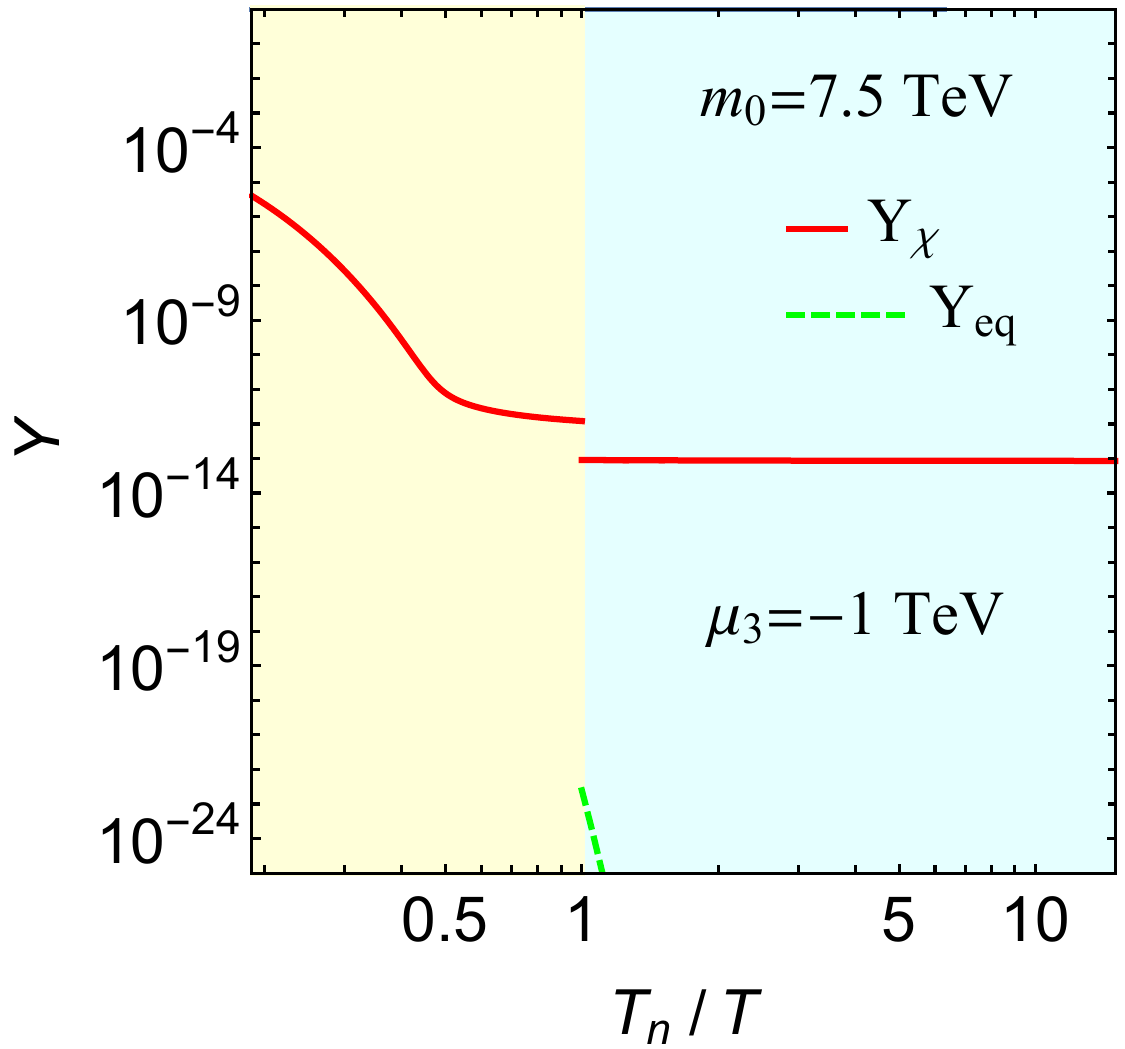}
	\caption{
DM freeze-out before the phase transition where $m_0=7.5$ TeV. The red line stands for  $Y_{\chi}$ and the green line stands for $Y_{\rm eq}$. The yellow(cyan) regime represents the case of outside(inside) the bubble, where the bubble nucleation temperature is located at $T_n =150~ {\rm GeV}.$}
\label{M}
\end{figure}

\subsection{Scenario B: DM freeze-out before the phase transition}

In this subsection we evaluate the penetration rate for the case where DM freezes out before the first order phase transition.  In this case $n_\chi^{\rm out}$ can be derived from solving the Boltzmann equation, but there is a problem in calculating $n_\chi^{\rm in}$ with  Eq.~(\ref{xxxx}) as we don't know the DM distribution function at the time of phase transition. We use the method developed in Ref.~\cite{Baker:2019ndr} and assume that the distribution function equals to the Boson-Einstein  distribution $f_\chi^{\rm eq}$ times a pre-factor $A(z,~p_z)$ which describes the deviation from equilibrium near the bubble wall
\begin{eqnarray}
f_\chi=A(z,p_z)f_\chi^{\rm eq}
\label{feq}
\end{eqnarray}
Then by employing the general Boltzmann equation $\textbf{L}[f_\chi]=\textbf{C}[f_\chi]$, where \textbf{L} is the Liouville operator and $\textbf{C}$ is the collision term, one can get the following partial differential equation with initial condition~\cite{Baker:2019ndr}
\begin{align}
a(z,p_z)\frac{\partial A(z,p_z)}{\partial z}+b(z)\frac{\partial A(z,p_z)}{\partial p_z}=& c(A,z,p_z),\nonumber \\
A\left( z\sim-\mathcal{O}(l_w), p_z>\sqrt{\Delta m^2}\right) =& A_0.
\label{PDE}
\end{align}
Here $l_w$ is the bubble wall thickness and $A_0$ satisfies
\begin{eqnarray}
A_0=n_\chi\left/\int \!\! \frac{d^3\textbf{p}}{(2\pi)^3}f_\chi^{\rm eq}(\textbf{p})\right.,
\end{eqnarray}
where $n_\chi$ is the DM number density  outside the bubble at $T \sim T_n$. It can be calculated by the Eq.~(\ref{beq}). Other terms in Eq.~(\ref{PDE}) are
\begin{align}
a(z,p_z)=& \frac{p_z}{m_\chi(z)}, ~~~~~~
b(z)= -\frac{\partial m_\chi(z)}{\partial z}, 
\end{align}
with
\begin{align}
m_\chi(z)=\sqrt{2\mu_2^2-\frac{9}{2\sqrt{2}}\mu_3v_s(z)},~~~~~~ v_s(z) = \frac{1}{2}v_s\left[1 + \tanh\left({3z\over l_w}\right)\right]
\end{align}
where $v_s=\left \langle S \right \rangle|_{T=0}$. The term $c(A,z,p_z)$ can be written as
\begin{align}
c(A,z,p_z)=& \frac{\partial m_\chi}{\partial z}\frac{v_w}{T_n}A(z,p_z)+ \frac{2\pi}{m_\chi T_n} {\rm exp}\left[\frac{\sqrt{m_{\chi}^2+p_{z}^2}-v_w p_z}{T_n}\right]\int \!\! \frac{dp_xdp_y}{(2\pi)^2}\textbf{C}\left[ f_\chi\right] .
%cf(z,p_z)=&\end{align}
\end{align}
The collision term $\textbf{C}[f_\chi]$ reads
\begin{align}
%g_\chi \int \! \frac{dp_x dp_y}{(2\pi)^2} 
\mathbf{C}[f_\chi]
&= \frac{1-A^2(z,p_z)}{2E_{\chi}}\int \!\!\frac{d^3\textbf{p}}{2E_{\chi}(2\pi)^3} \,
4 F \sigma_{\chi \chi \to SS}
 (f_{\chi}^\mathrm{eq})^2    
 \end{align}
where $F=\sqrt{s(s-4m_\chi^2)}/2$. 

\begin{figure}[htbp]
	\centering	\includegraphics[height=3in,angle=0]{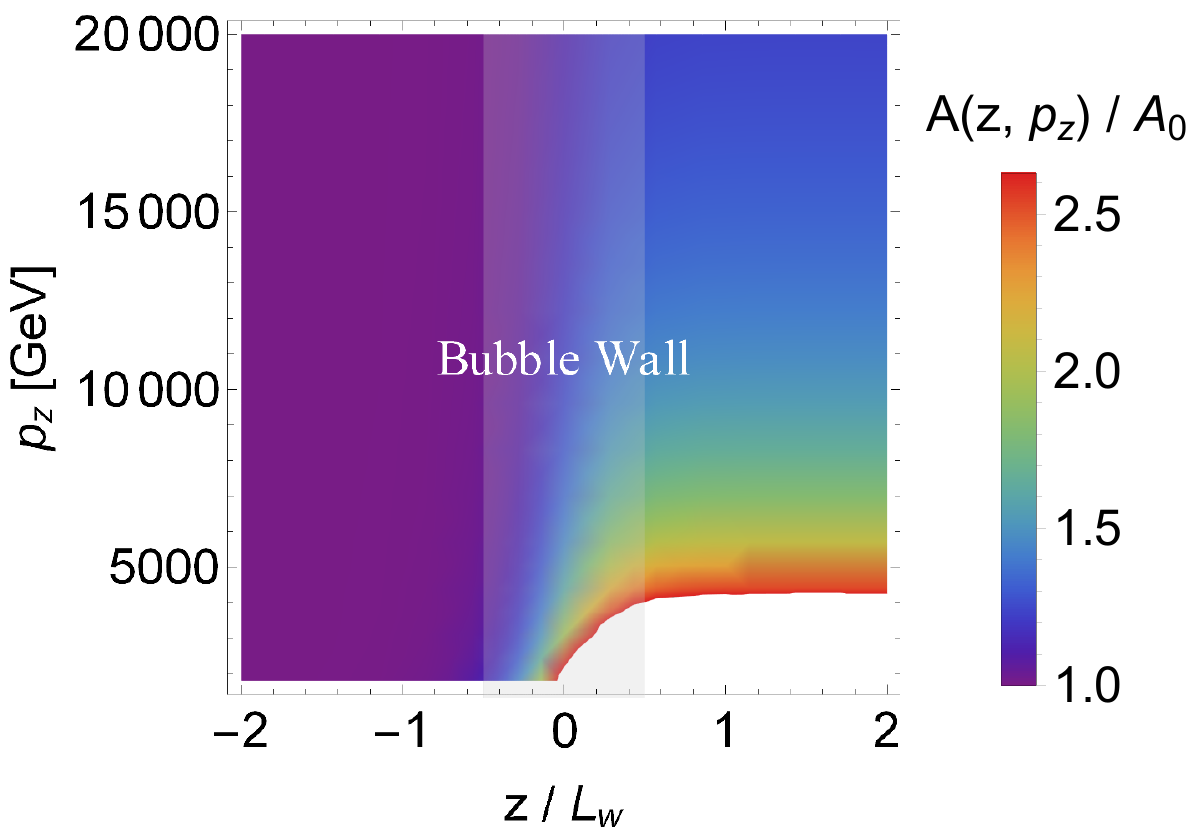}
	\caption{\small 
		Pre-factor $A(z,p_z)$ normalized by $A_0$ near the bubble wall. The gray band stands for the bubble wall and the white region still has $\mathcal{O}(A(z,p_z)/A_0)\sim 1$}
	\label{A}
\end{figure}

After calculating $A(z,p_z)$, we use Eq.~(\ref{feq}) to evaluate the DM number density $n_\chi^{\rm in}$ inside the bubble and then  yield $Y_\chi$ using Eq.~(\ref{beq}). We show in the Fig.~\ref{M}, $Y_\chi$ as the function of $T_n/T$ by setting $m_0=7.5~{\rm TeV}$ and $\mu_3=-1~{\rm TeV}$. As can be seen from the plot, DM freezes out before the phase transition. The penetration rate for this case  is $R_\chi \sim 7.41\times10^{-2}$.  In Fig.~\ref{A}, we show the pre-factor $A(z,p_z)$ near the bubble wall. 

\begin{figure}[htbp]
	\centering
	\includegraphics[height=1.8in,angle=0]{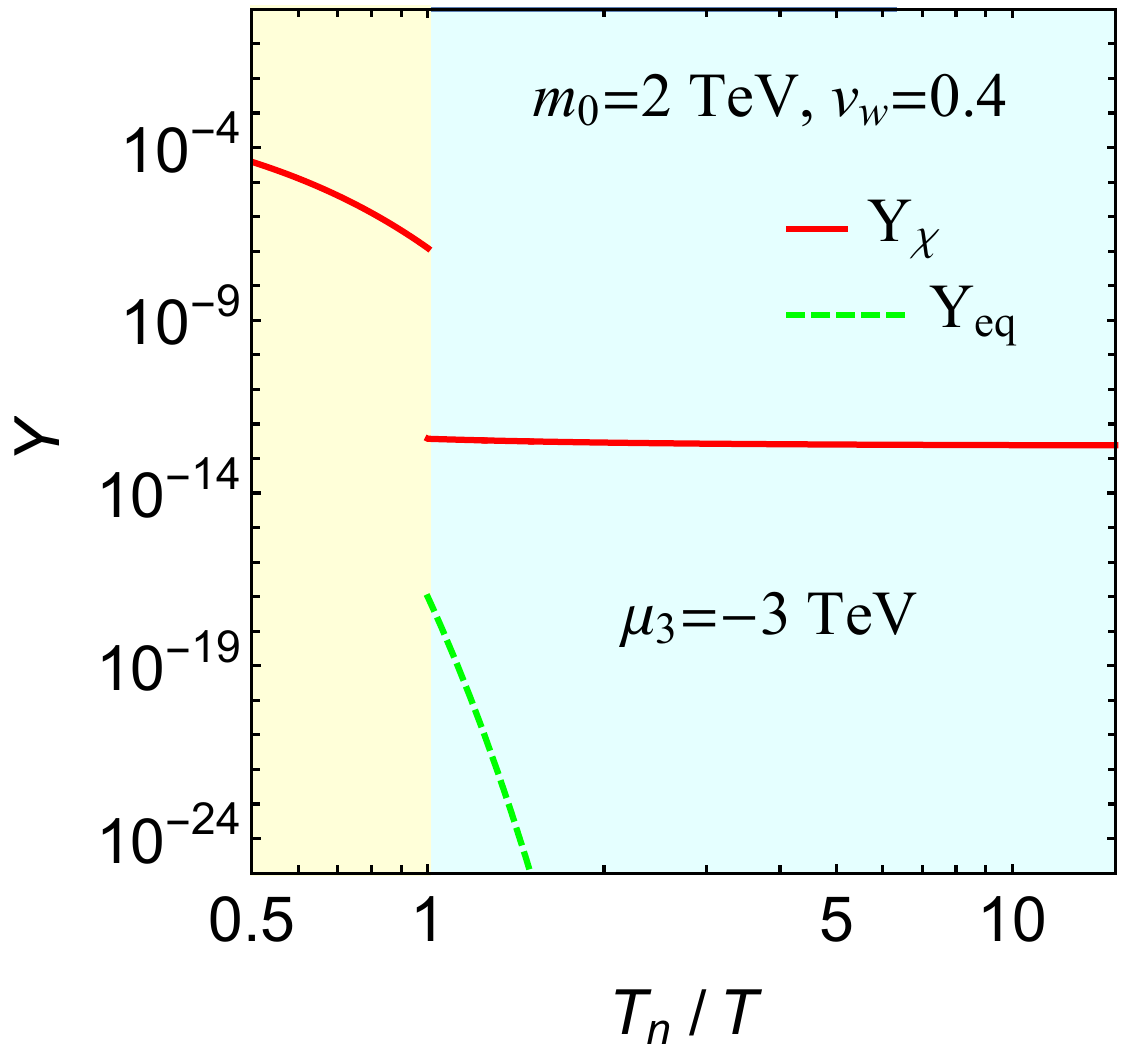}
	\includegraphics[height=1.8in,angle=0]{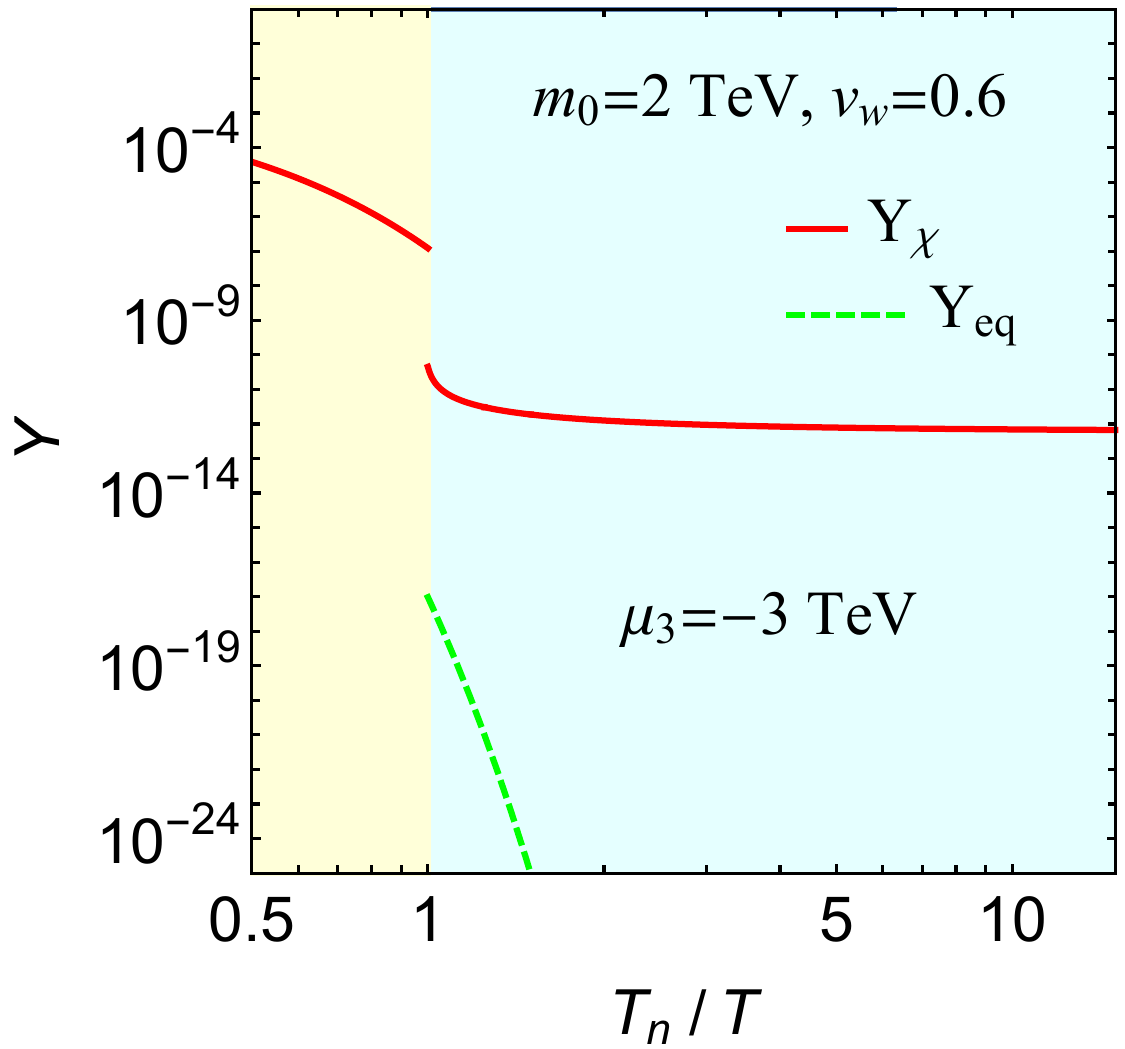}
	\includegraphics[height=1.8in,angle=0]{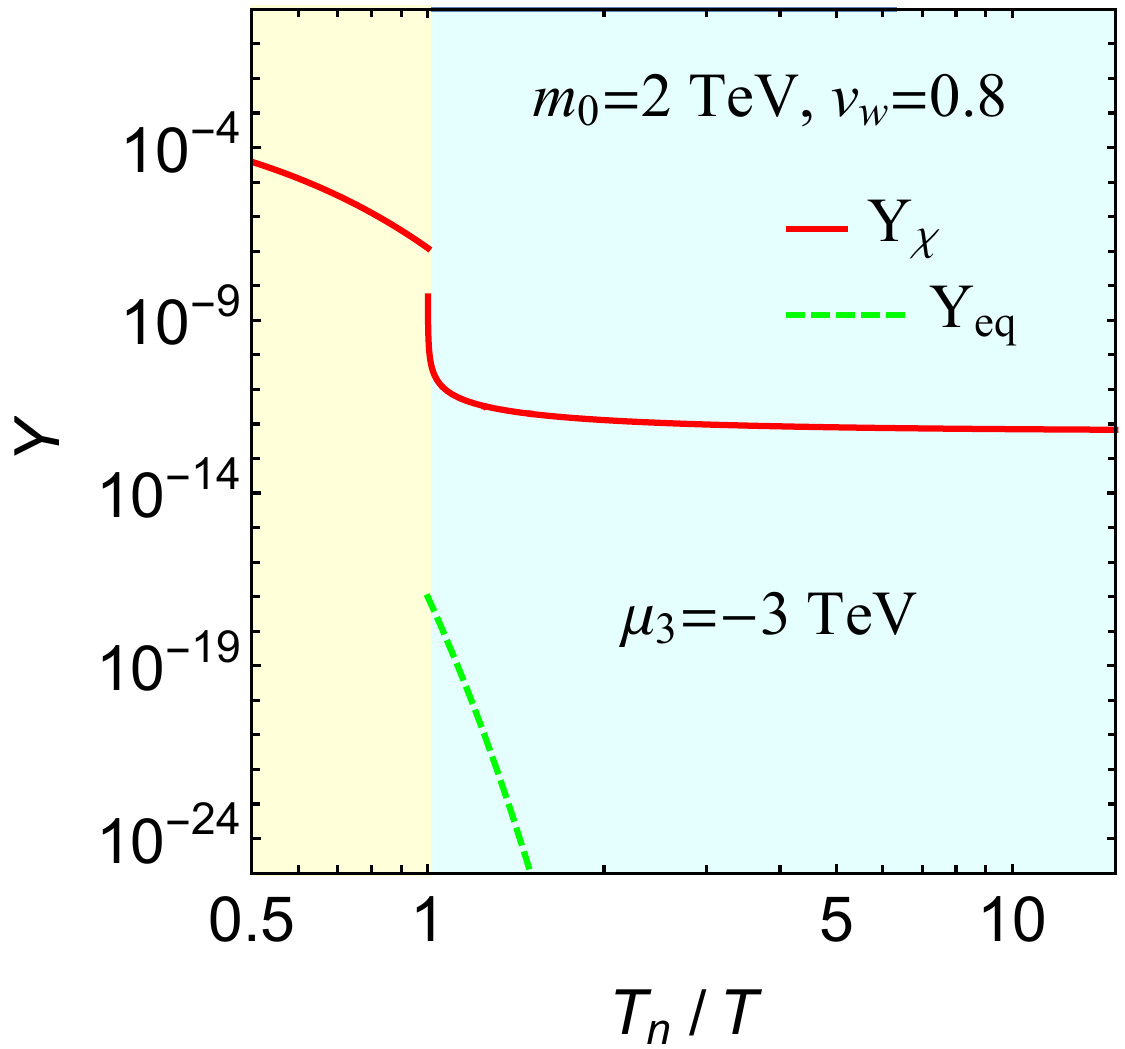}
\caption{ $Y$ as the function of $T_n/T$ with different bubble wall velocity $v_w$ in the case of $m_0=2$ TeV. The red line stands for  $Y_{\chi}$ and the green line stands for $Y_{\rm eq}$. The yellow(cyan) regime represents the case of outside(inside) the bubble, where the bubble nucleation temperature is located at $T_n =150~ {\rm GeV}.$
	}
	\label{v}
\end{figure}

\begin{figure}[htbp]
	\centering
	\includegraphics[height=2.7in,angle=0]{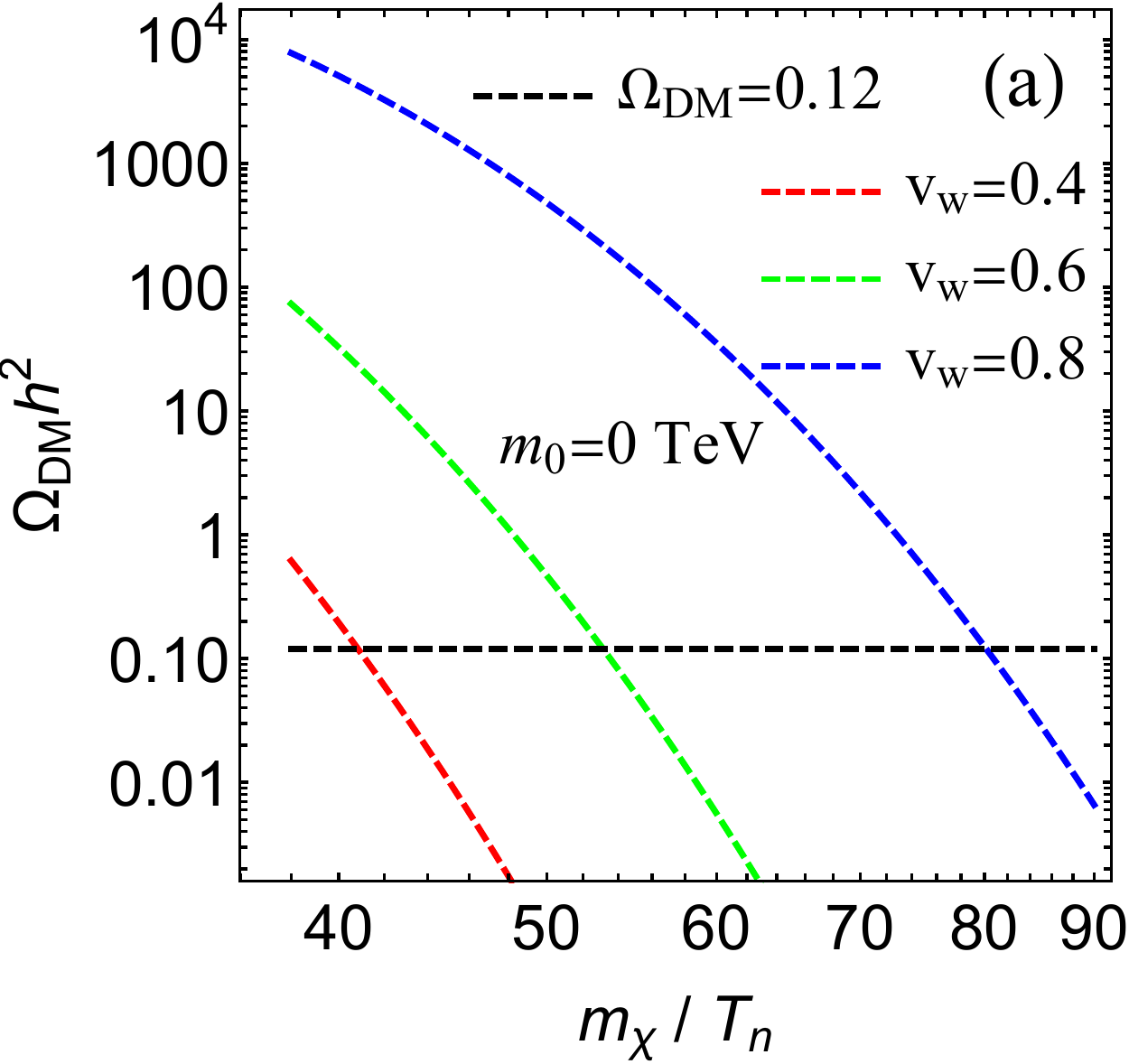}~~
	\includegraphics[height=2.7in,angle=0]{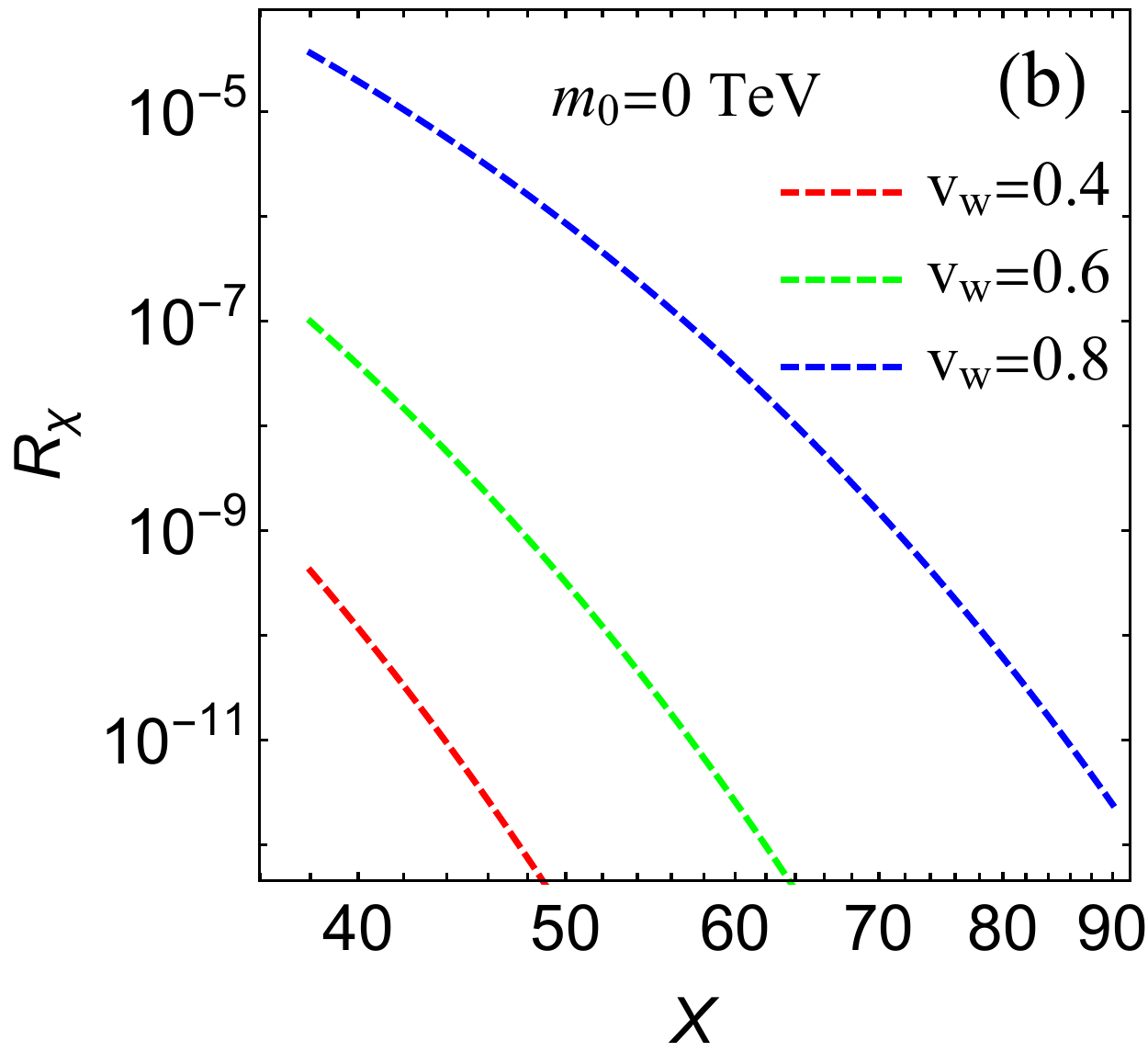} ~ \\
	\includegraphics[height=2.7in,angle=0]{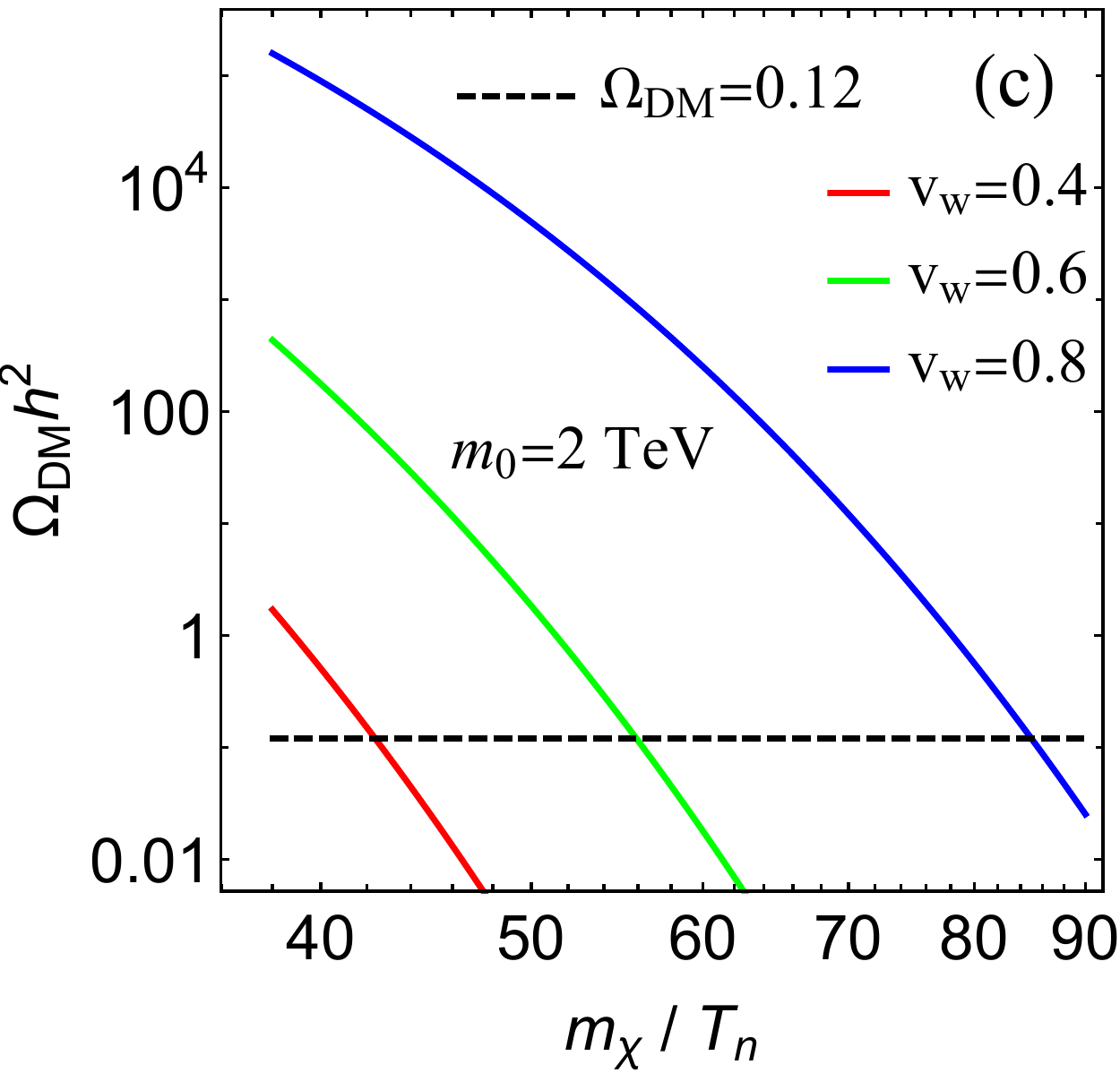}~~
	\includegraphics[height=2.7in,angle=0]{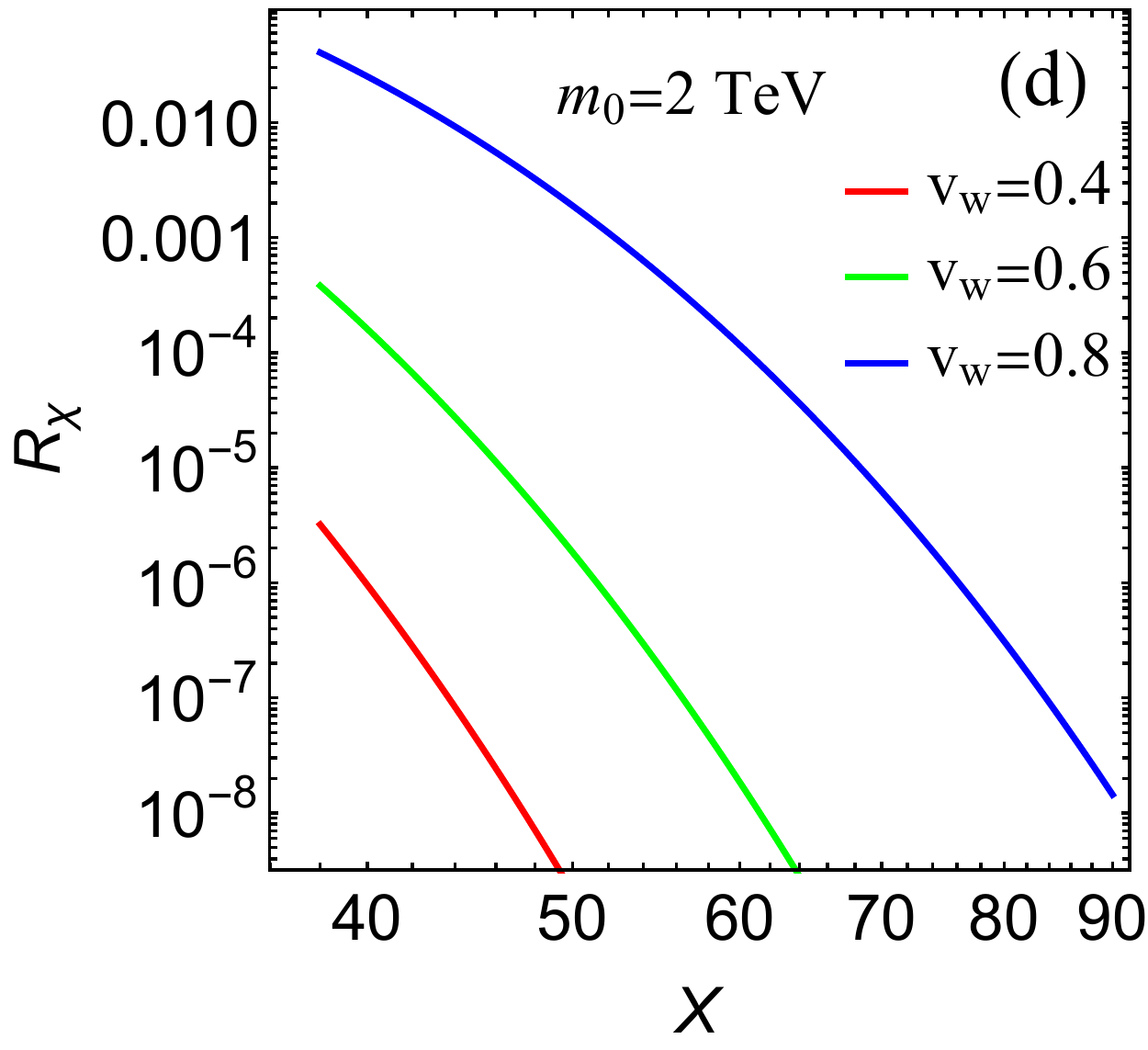}
	\caption{\small \label{fig:DM_relic}
		DM relic density $\Omega_{\rm DM} h^2$ and penetration rate $R_\chi$ with different $v_w$ and $M_{\chi}$ in the case of $T_n= 150~{\rm GeV}$. The red, green and blue lines represent $v_w=0.4,~0.6$ and $0.8$, respectively. Dashed lines (a)-(b) correspond to $m_0=0$ TeV and Solid lines (c)-(d) correspond to $m_0=2$ TeV. Black dashed line is the $\Omega_{\rm DM} h^2\sim0.12$.
	}
	\label{relic_v}
\end{figure}

%%%%%%%%%%%%%%% ????????  %%%%%%%%%%%%% 
%%%%%%%%%%%%%%% ????????  %%%%%%%%%%%%% 
%\begin{figure}[htbp]
%	\centering
	%\includegraphics[height=1.5in,angle=0]{relic_v0.2}
%	\includegraphics[height=2.25in,angle=0]{figures/n_v0.4}
%	\includegraphics[height=2.25in,angle=0]{figures/n_v0.8}
%	\caption{\small \label{fig:DM_relic}
%		The DM number density with different bubble wall velocity and DM mass before penetrating. Left $v_w=0.4$, Right $v_w=0.8$.  
%	}
%\end{figure}
%%%%%%%%%%%%%%%%%%%%%%%%%%%%%%%%%%%%%%%%%%%%% 
%%%%%%%%%%%%%%%%%%%%%%%%%%%%%%%%%%%%%%%%%%%%%

\subsection{Relic abundance with different bubble wall velocity $v_w$}
\label{3.4}

Finally let's evaluate the impacts of bubble wall velocity on the penetration rate. Dividing DM  energy density at $T_n$ by the entropy density $s=(2\pi^2/45)g_{\star S}T^3$ and normalizing to the critical density, $\rho_c = 3H^2_0 m^2_{\rm pl}$ where $m_{\rm pl}$ is the reduced Planck mass,  $H_0 = 100 \, h \ {\rm km} / \text{sec} / \text{Mpc}$ is the Hubble constant, $g_{\star S} \sim 106.75$ at $T_n$ and $g_{\star S}\equiv g_{\star S0}\sim 3.9$  today, we can get the current DM relic abundance~\cite{Marfatia:2020bcs,Baker:2019ndr}
\begin{align}
\label{eq:rescale}
\Omega_{\rm DM}h^2
&\simeq 
6.29\times 10^{8}\, 
\frac{m_\chi n^{\rm FO}_\chi}{\rm GeV} \frac{1}{g_{\star S}T^3_n},
%{\\ \notag
%&\simeq 1.595\times 10^7 
%\left( \frac{m_\chi}{\rm GeV} \right)
%\left( \frac{g_{\chi}}{g_{\star S}} \right)
%\left( \frac{1}{v_w} \right)
%\left( \frac{m_\chi}{T_n}+1 \right) 
\end{align}
where $n^{\rm FO}_\chi = n^{\rm in}_\chi$ for cases DM freezing-out during or before the phase transition.

In Fig.~\ref{v}, we show the impacts of bubble wall velocity $v_w$  to the penetration rate for the case $m_0=2$ TeV and $-\mu_3=3$ TeV. It is clear that the DM penetration rate is increasing with the increase of bubble wall velocity $v_w$. Next, we study the DM relic density $\Omega_{\rm DM}h^2$ for different bubble velocity $v_w$ at $T_n=150$ GeV in the cases of $m_0=0$ TeV and $m_0=2~{\rm TeV}$.  Fig.~\ref{relic_v} (a) and (c) show the $ \Omega_{\rm DM}h^2$ as the function of $m_\chi/T_n$ for $v_w=0.4,~0.6,~0.8$ with $m_0=0$ TeV and $m_0=2$ TeV, respectively.  Fig.~\ref{relic_v} (b) and (d) are their corresponding DM penetration rate $R_\chi$. From these plots, we can conclude that even if bubble velocity is big enough e.g. $v_w=0.8$, the bubble filtering-out effect on DM  may still be strong, and we can obtain the right DM relic density $\Omega_{\rm DM}h^2\sim0.12$ by taking a heavier DM mass. Therefore, it is possible  to detect the GW signals generated during the first order phase transition in the future space-based interferometer, which will be discussed in Sec.~\ref{4}.
%Fig.~\ref{relic2} shows the contour plot for DM relic density $\Omega_{\rm DM} h^2\left( m_0, -\mu_3\right)$ with $v_w=0.4$ and $0.8$. For $m_0=2000$ GeV and $v_w=0.4$, we fix $v_s=3200$ GeV and the right DM relic density $\Omega_{\rm DM}h^2\sim0.12$ corresponds to $-\mu_3\sim2000$ GeV. For $m_0=2000$ GeV and $v_w=0.8$, we fix  $v_s=6500$ GeV and the $\Omega_{\rm DM}h^2\sim0.12$  corresponds to $-\mu_3>7000$ GeV.
 
\begin{comment}
%%%%%%%%%%%%%%% ????????  %%%%%%%%%%%%% 
%%%%%%%%%%%%%%% ????????  %%%%%%%%%%%%% 
\begin{figure}[htbp]
	\centering	\includegraphics[height=2.3in,angle=0]{figures/relic_v04}~
	\includegraphics[height=2.3in,angle=0]{figures/relic_v08}\caption{\small 
	%	The DM relic density $\Omega_{\rm DM} h^2$ as a function of model parameters $\mu_3$ and $m_0=\sqrt{2}\mu_2$.
	Contour plot for DM relic density $\Omega_{\rm DM} h^2\left( m_0, -\mu_3\right) $ where $m_0=\sqrt{2}\mu_2$. The Solid lines stand for $\Omega_{\rm DM} h^2=0.1265$, and dashed lines represent $\Omega_{\rm DM} h^2=0.1133$.}
	\label{relic2}
\end{figure}
\end{comment}
%%%%%%%%%%%%%%%%%%%%%%%%%%%%%%%%%%%%%%%%%%%%%%%
%%%%%%%%%%%%%%%%%%%%%%%%%%%%%%%%%%%%%%%%%%%%%%%
%%%%%%%%%%%%%%%%%%%%%%%%%%%%%%%%%%%%%%%%%%%%%%%
%%%%%%%%%%%%%%%%%%%%%%%%%%%%%%%%%%%%%%%%%%%%%%%
\section{Stochastic gravitational waves}\label{4}
%%%%%%%%%%%%%%%%%%%%%%%%%%%%%%%%%%%%%%%%%%%%%%%%%%%%%%%%%%%%%%%%%%%%%%%%%%%%%%%%%%%%%%%%%%%%%%%%%%%%%%%%%%%%%%%%%

As can be seen from the potential in the Eq.~(\ref{potential1}), the cubic term may trigger the strongly first order phase transition.   As the Universe cools down to the critical temperature, phase transition takes place by the bubble nucleation with the tunneling rate per unit volume, $\Gamma\sim A(T) e^{-S_3/T}$, where $S_3$ is the Euclidean action of the critical bubble, 
\begin{eqnarray}
S_3 =4\pi\int r^2 dr \left[ {1\over 2 } \left( d \vec \phi(r) \over dr \right)^2 + V(\vec \phi, T)\right]
\end{eqnarray}
The bubble nucleation takes place when a single bubble can be nucleated within one horizon volume, which implies $S_3(T_n)/T_n\approx 140$~\cite{Apreda:2001us}, where $T_n$ is defined as the bubble nucleation temperature. 

For a first order phase transition, there will be stochastic GW generated from three processes: bubble collisions, sound waves in the plasma and Magnetohydrodynamic (MHD) turbulence~\cite{Caprini:2015zlo,Cai:2017cbj,Weir:2017wfa}.  The total energy spectrum is approximately the sum of these three parts:
\begin{eqnarray}
\Omega_{\rm GW} h^2 \approx \Omega_{\rm col} h^2 + \Omega_{\rm sw} h^2 + \Omega_{\rm turb} h^2 \; .
\end{eqnarray}
 The energy spectrum depends on three parameters: the bubble wall velocity $v_w$,  $\alpha$ and $\beta$ with 
\begin{eqnarray}
\alpha =\left.{\Delta \rho \over \pi^2 g_\star T^4 /30}\right |_{T=T_n}, \hspace{2cm} \beta =\left. H_n T_n {d S_3/T \over d T}\right|_{T=T_n}
\end{eqnarray}
where $\Delta \rho$ is the energy released to the true vacuum, $g_\star$ is the number of degrees of freedom, $H_n$ is the Hubble constant at $T_n$. 

In the following, we discuss the energy spectrum of the GW arising from these three sources.   The energy spectrum from the bubble collision can be calculated  by using numerical simulations~\cite{Huber:2008hg} or analytical approximations~\cite{Jinno:2016vai}.  It is  the function of $k^{}_\phi$, which is the fraction of latent heat transferred to the scalar field gradient.  According to recent studies~\cite{Bodeker:2017cim,Hoeche:2020rsg}, interactions of particles with the bubble wall are accompanied by the emission of soft gauge bosons, and  the bubble can not run away due to the thermal pressure exerted against the wall. As a result, energy deposited in the scalar field is very tiny i.e. $\kappa_\phi \ll 1$, and $\Omega_{\rm col} h^2$ is negligible.  

The bulk motion of the fluid in form of sound wave produced after the bubble collision generates the stochastic GW, with the energy spectrum~\cite{Hindmarsh:2015qta},
\begin{equation}
\Omega_{\textrm{sw}}h^{2}=2.65\times10^{-6}\left( \frac{H_{n}}{\beta}\right)\left(\frac{\kappa_{v} \alpha}{1+\alpha} \right)^{2} \left( \frac{100}{g_{\ast}}\right)^{1/3}v_{w} \left(\frac{f}{f_{sw}} \right)^{3} \left( \frac{7}{4+3(f/f_{\textrm{sw}})^{2}} \right) ^{7/2} \ .
\end{equation}
where $\kappa_v$ is the fraction of latent heat transformed into the bulk motion of the fluid,  $f_{\text{sw}}$ is the peak frequency at present time redshifted from the one at the phase transition,
 \begin{equation}
f_{\textrm{sw}}=1.9\times10^{-5}\frac{1}{v_{w}}\left(\frac{\beta}{H_{n}} \right) \left( \frac{T_{n}}{100~\textrm{GeV}} \right) \left( \frac{g_{\ast}}{100}\right)^{1/6} \textrm{Hz} .
\end{equation}
$\kappa_v$ is the function of $\alpha$ and $v_w$, and the  following approximate formulas~\cite{Espinosa:2010hh} will be used in our analysis:
\begin{itemize}
  \item For small bubble wall velocity ($v_w \ll c_s$),
\beq
\kappa_v \simeq v_w^{6/5} 
\frac{6.9 \alpha}{1.36 - 0.037 \sqrt{\alpha} + \alpha}\ ,
\eeq
where $c_s=1/\sqrt{3}$ denotes the sound velocity. 
  \item  For the transition from subsonic to supersonic deflagrations ($v_w =
c_s$),
\beq
\kappa_v \simeq \frac{\alpha^{2/5}}{0.017+ (0.997 + 
\alpha)^{2/5} }\ .
\eeq
\item For very large bubble wall velocity ($v_w \to 1$),
\beq
\kappa_v \simeq \frac{\alpha}
{0.73 + 0.083 \sqrt{\alpha} + \alpha}\ .
\eeq
\end{itemize}
Recent studies show that the development of shocks and turbulence within one Hubble time~\cite{Hindmarsh:2019phv,Ellis:2020awk}, as well as the formation of metastable vacuum~\cite{Cutting:2019zws} can disrupt the GW generated from the sound wave, resulting a slightly weaker GW signal.  

The plasma  is fully ionized at the time of phase transition, and the resulting MHD turbulence is a  third source of  stochastic GW with the spectrum.
The generated GW spectrum can be written as~\cite{Caprini:2009yp,Binetruy:2012ze},
\begin{equation}
\Omega_{\textrm{turb}}h^{2}=3.35\times10^{-4}\left( \frac{H_{n}}{\beta}\right)\left(\frac{\kappa_{\rm turb} \alpha}{1+\alpha} \right)^{3/2} \left( \frac{100}{g_{\ast}}\right)^{1/3} v_{w}  \frac{(f/f_{\textrm{turb}})^{3}}{[1+(f/f_{\textrm{turb}})]^{11/3}(1+8\pi f/h_{\ast})} ,
\end{equation}
where a possible helical component~\cite{Kahniashvili:2008pe} is neglected,  $\kappa_{\text{turb}}$  is the fraction of latent heat transferred to MHD turbulence, and the peak frequency $f_{\rm turb}$ is,
\begin{equation}
f_{\textrm{turb}}=2.7\times10^{-5}\frac{1}{v_{w}}\left(\frac{\beta}{H_{n}} \right) \left( \frac{T_{n}}{100~\textrm{GeV}} \right) \left( \frac{g_{\ast}}{100}\right)^{1/6} \textrm{Hz} .
\end{equation}
The precise value of $\kappa_{\rm turb}$ is still unknown, but it can be  parametrized as $\kappa_{\text{turb}}\approx \epsilon \kappa_{v} $, with the numerical  factor $\epsilon$ varying roughly between $5 \sim 10\%$~\cite{Hindmarsh:2015qta}. 
We set $\epsilon = 0.1$ in  the numerical simulations.

\begin{table}
	\begin{footnotesize}
		\begin{tabular}{| c | c | c | c | c | c| c| c| c|c|}
			\hline
			$v_s$ (GeV)&$\mu_2$ (GeV) &$\mu_3$ (GeV)&~~ $\mu_4$~~&$m_s$ (GeV) &$m_{\chi}$ (GeV)& $T_n$ (GeV) &$~~\beta/H_n$~~&~~$\alpha$~~\\
			\hline
			%%%%%%%hh
			511.6 &$0$ & $-1000^2/v_s$  & 0 & 735.2 & 1783.8 & 150.6 & 232.7 & 0.48\\
			\hline
		\end{tabular}
	\end{footnotesize}
	\caption{Input and output parameters for the benchmark point.}
	\label{tabgrav}
\end{table}

%%%%%%%%%%%%%%%%%%%%%

\begin{figure}[htbp]
	\centering
	\vspace{-2.7cm}
	\includegraphics[height=5.2in,angle=0]{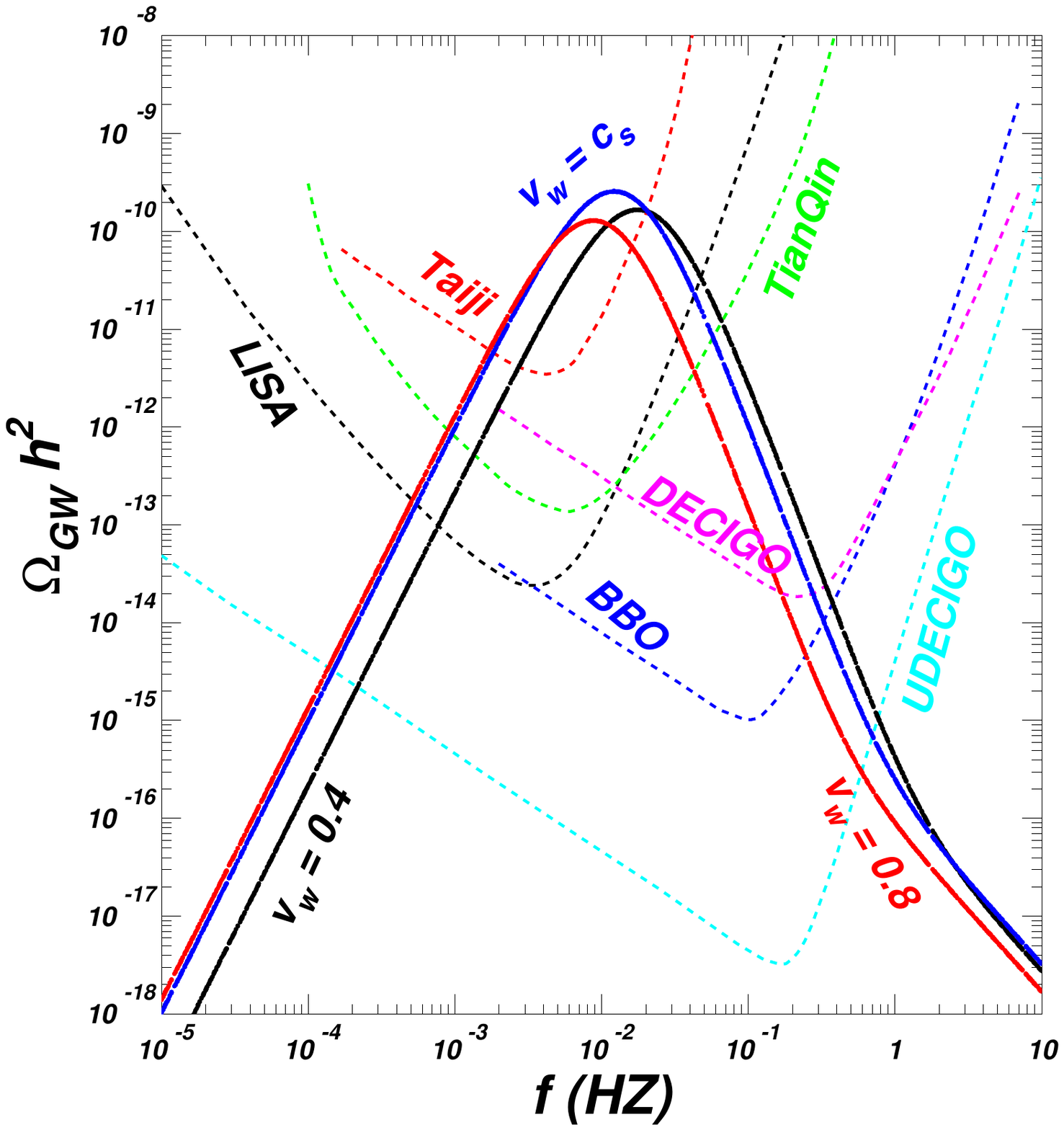}
	\vspace{-2.7cm}
	\caption{\small \label{2types}
		Gravitational wave spectra for the benchmark point.}
	\label{figgrav}
\end{figure}

We pick out a benchmark point  with relevant input and output parameters listed  in the Table \ref{tabgrav}, which correspond to inputs of the top-left plot in the Fig.~\ref{1types}, and present the details of the GW spectrum. The $\beta/H_n$ may characterize the inverse time duration of the phase transition.  A small $\beta/H_n$ means a long phase transition, and gives strong GW signals. In addition, a large $\beta/H_n$ can enhance the peak frequency of the GW spectrum. The parameter $\alpha$ describes the amount of energy released during the phase transition, and  a large $\alpha$ leads to strong GW signals. The phase transition for the benchmark point  leads a relatively large $\alpha$.

In Fig. \ref{figgrav}, we show the predicted GW spectrum for the benchmark point along with expected sensitivities of various future space-based interferometer experiments.  The black, blue and red solid lines represent the total energy spectrum for $v_w=0.4,~c_s$ and $0.8$, respectively.  The regions surrounded by dashed lines are the experimental sensitivity regions for the Taiji (magenta), LISA(black), TianQin(green), DECIGO(pink), BBO(blue) and Ultimate-DECIGO(cyan), respectively.  Due to the enhancement of a large $\alpha$, the amplitudes of the GW spectrum reach the sensitivities of all experiments.

%%%%%%%%%%%%%%%%%%%%%%%%%%%%%%%%%%%%%%%%%%%%%%%
%%%%%%%%%%%%%%%%%%%%%%%%%%%%%%%%%%%%%%%%%%%%%%%%%%%%%%%%%%%%%%%%%%%%%%%%%%%%%%%%%%%%%%%%%%%%
\section{Summary and outlook}\label{Sec_Numerical}

The thermal history of a DM is  worthy of deep study. In this paper we have investigated the freeze-out of  a pseudo-scalar DM with special focus on the filtering-out effect induced by the first order phase transition in the dark sector. We have calculated the penetration rate, which characterizes the filtering-out effect,  for three typical scenarios: the  DM freeze-out before, during and after the first order phase transition. We find that the filtering-out effect can be significant for heavy DM and this effect may  allow the existence of super-heavy DM with correct relic abundance without violating the unitarity.  We have calculated the spectrum of stochastic gravitational wave emitted during the first order phase transition as a smoking-gun of this scenario. Our results of a typical scenario show that these gravitational wave signal can be observed by the future space-based interferometer. The filtering-out effect for the fermionic DM have been delicately investigated in Refs.~\cite{Baker:2019ndr,Chway:2019kft}, our study is complementary to them by addressing the detail of a new scenario, in which DM  may get non-zero mass before the phase transition and may freeze-out outside the bubble.

%%%%%%%%%%%%%%%%%%%%%%%%%%%%%%%%%%%%%%%%%%%%%%%%%%%%%%%%%%%%%%%%%%%%%%%%%%%%%%%%%%%%%%%%%%%%%%%
\section*{Acknowledgments}
This work was supported by the National Natural Science Foundation of China under grant No. 11775025, No. 11975013 and the Fundamental Research Funds for the Central Universities under grant No. 2017NT17.
%\section*{References}

%%%%%%%%%%%%%%%%%%%%%%%%%%%%%%%%%%%%%%%%%%%%%%%%%%%%%%%%%%%%%%%%%%%%%%%%%%%%%%%%%%%%%%%%%%%%%
\appendix
\section{Effective potential including the Higgs portal interaction}\label{appA}
%%%%%%%%%%%%%%%%%%%%%%%%%%%%%%%%%%%%%%%%%%%%%%%%%%
A  general Higgs potential including the Higgs portal interaction has the form
\begin{eqnarray}
\label{potential2}
V(H,S) &=& -\mu_h^2 |H|^2 -\mu_s^2 |S|^2 + \lambda_h |H|^4 + \lambda_s |S|^4 + \lambda_{sh} |H|^2 |S|^2 \nonumber \\
&& - \frac{\mu_2^2}{2}\left(S^2 + {S^\star}^2 \right)\ + \frac{\mu_3}{2}\left(S^3 + {S^\star}^3 \right)+ \frac{\mu_4}{2}\left(S^4 + {S^\star}^4 \right),
\label{tree}
\end{eqnarray}
where $H$ denotes the SM Higgs doublet. The Higgs and singlet scalar can be expanded around their classical backgrounds as
\begin{eqnarray}
H=\left(\begin{array}{c}{G^{+}} \\ {\frac{1}{\sqrt{2}}\left(v_h+h+i G^{0}\right)}\end{array}\right),~~S=\frac{v_s+s+i\chi}{\sqrt{2}},
\end{eqnarray}
where $G^{\pm}$, $G^{0}$, and $\chi$ are the Goldstone bosons after spontaneous symmetry breaking. Eq.~(\ref{potential1}) is invariant under $\chi \to -\chi$, therefore $\chi$ is stable and can be considered as a DM candidate. At zero temperature, the vacuum expectation
values (VEVs) for the two scalars are $v_h \equiv \left \langle H \right \rangle|_{T=0}$ and 
$v_s \equiv \left \langle S \right \rangle|_{T=0}$, respectively. 
Then by minimizing the scalar potential, we obtain the following two conditions
\begin{eqnarray}
\label{mini1}
v_h(\mu_h^2 - \lambda_h v_h^2 - \frac{1}{2}\lambda_{sh} v_s^2) = 0,\\
\label{mini2}
v_s(\mu_s^2  -\lambda_s v_s^2 -\frac{1}{2}\lambda_{sh} v_h^2+\mu_2^2 -\frac{3}{2\sqrt{2}}\mu_3v_s- \mu_4 v_s^2) = 0.
\end{eqnarray}
The field-dependent mass matrix of the scalar bosons is given by
\begin{align}
\begin{split}
\mathcal{M}^2 =&
\begin{pmatrix}
\mathcal{M}^2_{hh} && \mathcal{M}^2_{hs}\\ 
\mathcal{M}^2_{sh} &&  \mathcal{M}^2_{ss}
\end{pmatrix}=
\begin{pmatrix}
2\lambda_h v_h^2 &&& \lambda_{sh} v_h v_s\\ 
\lambda_{sh} v_h v_s &&&  2\lambda_s v_s^2 +\frac{3}{2\sqrt{2}}\mu_3 v_s + 2\mu_4 v_s^2 
\end{pmatrix}.
\end{split}
\label{eq:mass1}
\end{align} 
The mass matrix  Eq.~(\ref{eq:mass1}) is diagonalised by an orthogonal
matrix
\begin{align}
\label{matrix}
\mathcal{O}=
\begin{pmatrix}
\cos\theta  & \sin\theta  \\ 
-\sin\theta  & \cos\theta 
\end{pmatrix},
\end{align}
via diag($m_{h}^2,m_{s}^2)=\mathcal{O^T} \mathcal{M}^2\mathcal{O}$ with the mixing angle $\theta$ is
\begin{eqnarray}
\label{angle}
\tan 2\theta =\frac{\lambda_{sh} v_h v_s}{\lambda_s v_s^2 + \frac{3}{4\sqrt{2}}\mu_3 v_s +\mu_4 v_s^2 -\lambda_h v_h^2 }.
\end{eqnarray} 
The field-dependent mass of pseudo-Goldstone boson $\chi$ is
\begin{equation}
\label{mchi}
m_{\chi}^2=-\mu_s^2+ \lambda_s v_s^2 +\frac{1}{2}\lambda_{sh} v_h^2 +\mu_2^2 -\frac{3}{\sqrt{2}}\mu_{3} v_s- 3\mu_4 v_s^2= 2\mu_2^2 -\frac{9}{2\sqrt{2}}\mu_3 v_s -4\mu_4 v_s^2.
\end{equation} 
Then we can express $\lambda_h$, $\lambda_s$,  $\lambda_{sh}$, $\mu_h^2$, $\mu_s^2$  with freedom quantities $m_{h},m_{s},m_{\mathcal{\chi}}, v_h,v_s$, $\mu_3$,  $\mu_4$ and $\theta$
\begin{align}
\label{m}
\lambda_h=& \frac{m_{h}^2  +m_{s}^2+ (m_{h}^2 -m_{s}^2)\cos 2\theta}{4 v_h^2},\\  
%(a^2 + a^2 f + g^2 - f g^2)/4 b^2
%\frac{M_{\mathcal{H}}^2 +M_{\mathcal{S}}^2 +(M_{\mathcal{H}}^2 -M_{\mathcal{S}}^2)\cos 2\theta}{4 v_h^2},\\
\lambda_s=& \frac{ m_{h}^2+  m_{s}^2+ (m_{s}^2 -m_{h}^2)\cos 2\theta -\frac{3}{\sqrt{2}}\mu_3 v_s  - 4\mu_4 v_s^2  }{4 v_s^2},\label{lambdas}\\
%\frac{3(M_{\mathcal{H}}^2 +M_{\mathcal{S}}^2)-2m_\chi^2 +3(M_{\mathcal{S}}^2 -M_{\mathcal{H}}^2)\cos 2\theta}{12 v_s^2},\\
\lambda_{sh}=&\frac{(m_{s}^2 -m_{h}^2)\sin 2\theta}{2 v_h v_s},\\
\mu_h^2=&\frac{m_{h}^2+ m_{s}^2}{4}+ \frac{(m_{h}^2- m_{s}^2)(v_h \cos 2\theta -v_s \sin 2\theta)}{4v_h},\\
%%%\mu_s^2=&\frac{M_{\mathcal{H}}^2+ M_{\mathcal{S}}^2-4\mu_2^2+\frac{3}{\sqrt{2}}\mu_3v_s}{4}+ \frac{(M_{\mathcal{S}}^2- M_{\mathcal{H}}^2)(v_h \sin 2\theta +v_s \cos 2\theta)}{4v_s},\label{mus}\\
%%%\mu_2^2=& \frac{m_\chi^2+ \frac{9}{2\sqrt{2}}\mu_3 v_s {\textcolor{red}{+}}4\mu_4 v_s^2}{2}. 
\mu_s^2=&\frac{m_{h}^2+ m_{s}^2- 2m_\chi^2-\frac{6}{\sqrt{2}}\mu_3 v_s- 8\mu_4 v_s^2}{4}+ \frac{(m_{s}^2- m_{h}^2)(v_h \sin 2\theta +v_s \cos 2\theta)}{4v_s},
\label{mus2} \\
\mu_2^2=& \frac{m_\chi^2+ \frac{9}{2\sqrt{2}}\mu_3 v_s + 4\mu_4 v_s^2}{2}. 
\end{align}
%%%%%%%%%%%%%%%%%%%%%%%%%%%%%%%%%%%%%%%%%%%%%%%%%%

The effective potential at finite temperature includes the tree level  potential, the Coleman-Weinberg term \cite{Coleman:1973jx}, the finite temperature corrections \cite{Dolan:1973qd,Quiros:1999jp} and the daisy resummation \cite{Gross:1980br,Parwani:1991gq}, which is gauge-dependent \cite{Jackiw:1974cv,Patel:2011th}.  Here we can take a gauge invariant approxmation, which keeps only the thermal mass terms in the high-temperature expansion. This effective potential is then given by
\begin{eqnarray}
\label{eq:cterm}
V_{\rm eff}(h, s, T)&=&
-\frac{1}{2} [\mu_{h}^{2}- \Pi_h(T)] h^2
-\frac{1}{2}[\mu_{s}^{2}+ \mu_2^2- \Pi_s(T)] s^2
\nonumber \\
&&+\frac{1}{4} \lambda_{h} h^{4}
+{1\over 4}(\lambda_{s}+ \mu_4) s^{4}
+\frac{1}{4} \lambda_{sh} h^{2} s^{2}
%-\mu_{2}^{2} s^{2}
+\frac{1}{2\sqrt{2}}\mu_{3} s^{3}
%+\mu_{4} s^{4}
\end{eqnarray}
where $\Pi_h(T)$ and $\Pi_s(T)$ are the thermal masses of the scalar fields \cite{Huang:2020bbe}
\begin{eqnarray}
\label{pi_h}
\Pi_h(T)&=&\frac{T^2}{48}(
9 g^{2}+3 g^{\prime 2}+12y_{t}^{2}+4\lambda_{sh}+24\lambda_{h}),\\
\label{pi_s}
\Pi_s(T)&=&\frac{T^2}{3}\left(\frac{1}{2}\lambda_{sh}+\lambda_{s} \right),
\end{eqnarray}
where $g$ and $g^{\prime}$ are the SM gauge couplings of $SU(2)_{L}$ and $U(1)_{Y}$, $y_t$ is the top quark Yukawa coupling.

For a simplified case where $\mu_4=0$ and only $S$ is included , the effective potential is 
\begin{equation}
V_{\rm eff}(s,T)=
%-\frac{1}{2} \mu_{h}^{2} h^{2}
%-\frac{1}{2}\mu_{s}^{2} s^{2}
%+\frac{1}{4}\lambda_{h} h^{4}
%+ \frac{1}{4}\lambda_{s} s^{4}
%+\frac{1}{4} \lambda_{hs} h^{2} s^{2}
%-\frac{1}{2}\mu_{2}^{2} s^{2}
%+\frac{1}{2\sqrt{2}}\mu_{3} s^{3}
%+\frac{1}{4}\mu_{4} s^{4}.
%-\frac{1}{2} [\mu_{h}^{2}- \Pi_h(T)] h^2
-\frac{1}{2} \left( \mu_{s}^{2}+ \mu_2^2- \lambda_s\frac{T^2}{3}\right)  s^2
%%\nonumber \\
%%&&+\frac{1}{4} \lambda_{h} h^{4}
+{1\over 4} \lambda_{s} s^{4}
%%+\frac{1}{4} \lambda_{hs} h^{2} s^{2}
%-\mu_{2}^{2} s^{2}
+\frac{1}{2\sqrt{2}}\mu_{3} s^{3}.
%+\mu_{4} s^{4}
\label{potential}
\end{equation}
Then the critical temperature $T_c$ and critical field value $v_s(T_c)$ in broken phase can be obtained by
\begin{align}
V_{\rm eff}(0,T_c)= V_{\rm eff}\left(v_s(T_c),T_c\right),~~~~~ \frac{\partial V_{\rm eff}(s,T_c) }{\partial{s}}|_{s=v_s(T_c)}=0.
\end{align}
\begin{comment}
It is obvious that
\begin{align}
T_c=\frac{\sqrt{12\lambda_s\left(\mu_2^2+\mu_s^2\right)+3\mu_3^2}}{2\lambda_s},~~~~~ v_s|_{T=T_c}=\frac{-\mu_3}{\sqrt{2}\lambda_s}~~(\rm for~  \mu_3<0).
\end{align}
\end{comment}
To ensure the tree-level part in Eq.~(\ref{potential}) is bounded from below and the mass matrix is definite, we have
\begin{eqnarray}
\lambda_s>0,~~~~~~8\lambda_s v_s +3\sqrt{2}\mu_3>0.
\label{det}
\end{eqnarray}
%%%%%%%%%%%%%%%%%%%%%%%%%%%%%%%%%%%%%%%%%%%%%%%%%%%
%Another method for calculating DM number density
%%%%%%%%%%%%%%%%%%%%%%%%%%%%%%%%%%%%%%%%%%%%%%%%%%%

%\begin{thebibliography}{0}
\bibliographystyle{JHEP}
\bibliography{filtering}

%\end{thebibliography}

\end{document}